\documentclass[reprint]{revtex4-2}

\usepackage{graphicx}
\usepackage{dcolumn}
\usepackage{bm}
\usepackage{float}
\usepackage{xcolor}
\usepackage{mhchem}
\usepackage{gensymb}
\usepackage{amsmath}
\usepackage{verbatim}

\begin{document}

\preprint{APS/123-QED}

\title{Positive temperature-dependent thermal conductivity induced by wavelike phonons in complex Ag-based argyrodites}

\author{Niuchang Ouyang\textsuperscript{1}}
\email{These authors contributed equally to this work}
\author{Dongyi Shen\textsuperscript{1}}
\email{These authors contributed equally to this work}
\author{Chen Wang\textsuperscript{2, 1}}
\email{These authors contributed equally to this work}
\author{Ruihuan Cheng\textsuperscript{1}}
\author{Qi Wang\textsuperscript{3}}
\author{and Yue Chen\textsuperscript{1}}
\email{yuechen@hku.hk}
\affiliation{\textsuperscript{1}Department of Mechanical Engineering, The University of Hong Kong, Pokfulam Road, Hong Kong SAR, China}
\affiliation{\textsuperscript{2}Institute for Advanced Study, Shenzhen University, Shenzhen 518060, China}
\affiliation{\textsuperscript{3}Thermal Science Research Center, Shandong Institute of Advanced Technology, Jinan, Shandong Province 250103, China}


\date{\today}

\begin{abstract}
The phonon transport mechanisms and the anomalous temperature-dependent lattice thermal conductivities ($\kappa_{\rm L}$) in Ag-based argyrodites have not been fully understood. 
Herein, we systematically study the phonon thermal transport of five Ag-based crystalline argyrodites $\rm Ag_{7}PS_{6}$, $\rm Ag_{7}AsS_{6}$, $\rm Ag_{8}SnS_{6}$, $\rm Ag_{8}GeS_{6}$ and $\rm Ag_{9}GaS_{6}$ utilizing perturbation theory and the unified theory thermal transport model. 
Our results show that, as the complexity of the unit cell increases, the proportion of the population terms declines while the coherence contributions become more significant, leading to the relfatively weak temperature-dependent $\kappa_{\rm L}$ of $\rm Ag_{7}PS_{6}$ and $\rm Ag_{7}AsS_{6}$, while the more complex crystalline argyrodites, $\rm Ag_{8}SnS_{6}$, $\rm Ag_{8}GeS_{6}$ and $\rm Ag_{9}GaS_{6}$, exhibiting a glass-like behavior in their temperature dependence of $\kappa_{\rm L}$.
We attribute the positive temperature-dependent and ultralow $\kappa_{\rm L}$ of $\rm Ag_{8}SnS_{6}$, $\rm Ag_{8}GeS_{6}$ and $\rm Ag_{9}GaS_{6}$ to the dominance of wavelike phonons and the strong phonon broadening.
Furthermore, using laser flash measurements and the homogeneous non-equilibrium molecular dynamics simulations based on accurate machine learning neuroevolution potentials, we provide further evidence for the glass-like temperature-dependent $\kappa_{\rm L}$ of $\rm Ag_{8}SnS_{6}$ and $\rm Ag_{8}GeS_{6}$. 
\end{abstract}

\maketitle

\section{INTRODUCTION}
Materials with extreme heat transport properties are of utmost technological importance, as they have a wide range of applications, including heat dissipation in electronic devices, thermal barrier coatings, and energy conversion in thermoelectric devices \cite{bell2008cooling, behnia2015fundamentals, zhang2015thermoelectric, humphrey2005reversible}. Research efforts in the field of thermoelectric materials have been directed toward minimizing irreversible heat transport with the objective of enhancing the efficacy of thermoelectric conversion \cite{snyder2008complex, sootsman2009new, tritt2011thermoelectric, snyder2003thermoelectric}. As a result, the study of heat transport through solids has become a prevalent technological topic. In simple crystals, the lattice thermal conductivity ($\kappa_{\rm L}$) can be well described by the phonon-gas model (PGM) \cite{mcgaughey2019phonon}. Generally, the PGM is universal in heat conductors that exhibit crystal-like behavior and can successfully explain the heat transport of some materials such as those with high $\kappa_{\rm L}$ \cite{dong2014size} because they have a regular crystal lattice that enables phonons to move freely through the system. However, the PGM is limited in describing the thermal transport behavior of some amorphous materials \cite{allen1999diffusons, isaeva2019modeling}, defective crystals \cite{seyf2017rethinking}, anharmonic materials \cite{ouyang2022role, ouyang2022temperature}, hybrid perovskites \cite{wang2023b, zheng2022anharmonicity}, and Zintl-type compounds \cite{hanus2021uncovering, wang2022intrinsic} with complex crystal structures due to their unique structural features and unconventional thermal transport properties. Simoncelli \textit{et al.} \cite{simoncelli2019unified} and Isaeva \textit{et al.} \cite{isaeva2019modeling} have recently developed two-channel lattice dynamics simulations, which are effective in quantitatively characterizing thermal transport in solids, encompassing both crystal-like and amorphous-like behaviors \cite{luo2020vibrational, xia2020particlelike, allen1989thermal}. Nevertheless, understanding the intricate physics that governs heat conduction in complex glass-like materials remains a significant challenge due to their complicated nature, which falls between typical crystals and glasses.

Herein, we focus on the phonon thermal properties of five Ag-based crystalline argyrodites, $\rm Ag_{7}PS_{6}$, $\rm Ag_{7}AsS_{6}$, $\rm Ag_{8}SnS_{6}$, $\rm Ag_{8}GeS_{6}$ and $\rm Ag_{9}GaS_{6}$, which have complex crystal structures consisting of over 56 atoms in their primitive unit cells.
Recently, they have attracted immense interest due to their remarkable thermoelectric properties and practical advantages, such as easy synthesis and the non-toxic, abundantly available elements in their compositions \cite{sturm2021stability, ghrib2015high, pogodin2022crystal, shen2023soft}. 
Although the thermodynamic and electronic transport properties of $\rm Ag_{7}PS_{6}$ and $\rm Ag_{7}AsS_{6}$ have been experimentally investigated \cite{pogodin2022crystal, carcaly1979composes}, a thorough understanding of the phonon thermal transport is still needed. Despite being ordered high-symmetry crystals, $\rm Ag_{8}SnS_{6}$, $\rm Ag_{8}GeS_{6}$ and $\rm Ag_{9}GaS_{6}$ display ultralow $\kappa_{\rm L}$ around 0.20 $\sim$ 0.34 $\rm Wm^{-1}K^{-1}$ at room temperature \cite{petrov1975characteristics, lin2018thermoelectric}. More interestingly, $\rm Ag_{8}SnS_{6}$ and $\rm Ag_{8}GeS_{6}$ exhibit a fairly positive temperature dependence of $\kappa_{\rm L}$ across a wide temperature range from 200 to 430 K \cite{petrov1975characteristics}, resembling that of amorphous materials, rather than the typical $T^{-1}$ dependence of crystalline materials \cite{feng2017four}. 
However, a further insight into the microscopic heat transfer mechanisms underlying ultralow and the positive temperature-dependent $\kappa_{\rm L}$ remains lacking. 

\begin{figure*}
\centering
\includegraphics[width=1\linewidth]{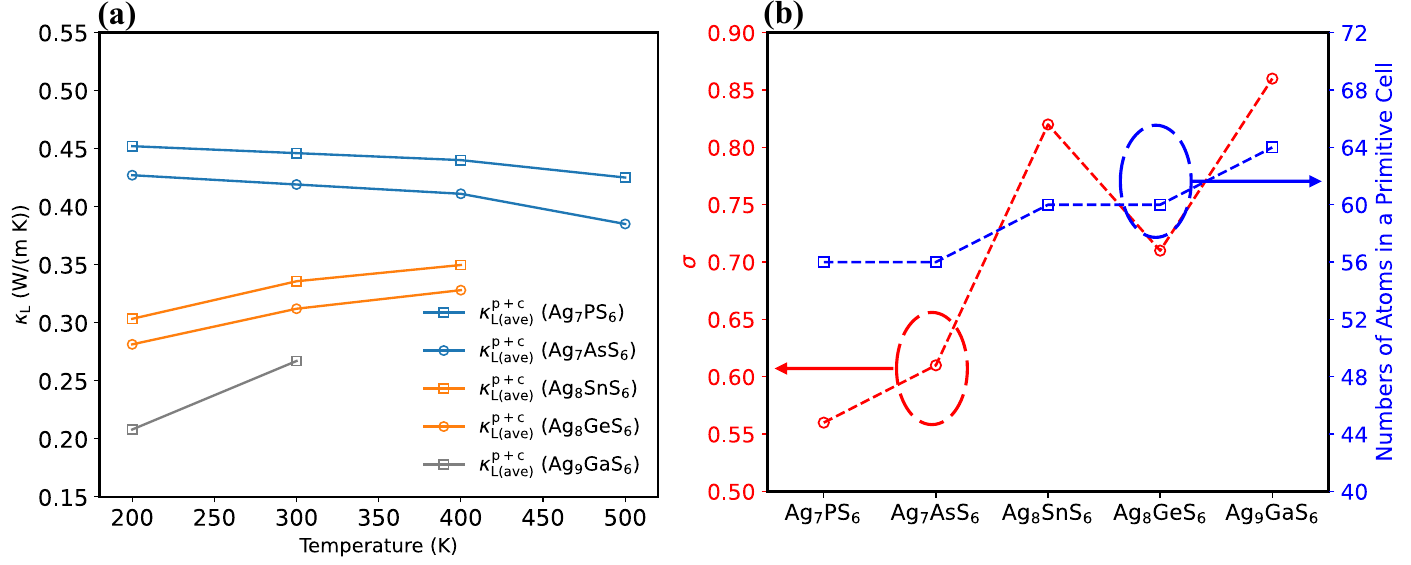}
\caption{\label{fig:wide} (a) Temperature-dependent $\kappa_{\rm L}$ of five Ag-based crystalline argyrodites, $\rm Ag_{7}PS_{6}$, $\rm Ag_{7}AsS_{6}$, $\rm Ag_{8}SnS_{6}$, $\rm Ag_{8}GeS_{6}$ and $\rm Ag_{9}GaS_{6}$ calculated within UT using the Phono3py \cite{togo2023first} package for the phonon cross group velocities. (b) Degree of anharmonicity $\sigma$ of argyrodites at 300 K and the number of atoms in their primitive cells. The average lattice thermal conductivity ($\kappa_{\rm L(ave)}$) is computed by taking the arithmetic average of $\kappa_{\rm L}$ along three Cartesian directions.}
\label{fig:1-1}
\end{figure*}


\begin{figure}
\centering
\includegraphics[width=1\linewidth]{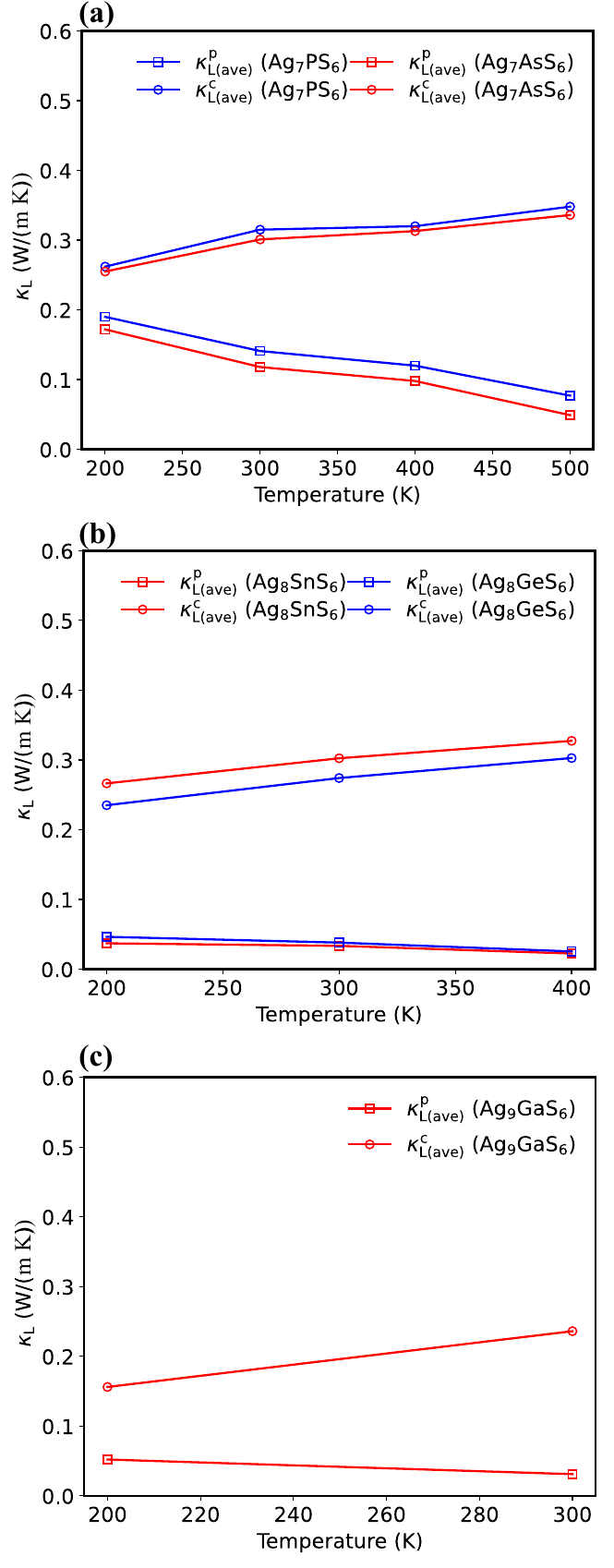}
\caption{\label{fig:wide} The phonon population and coherence contributions to $\kappa_{\rm L}$ of $\rm Ag_{7}XS_{6}$ (X=P, As) (a), $\rm Ag_{8}YS_{6}$ (Y=Sn, Ge) (b) and $\rm Ag_{9}GaS_{6}$ (c) calculated within UT using the Phono3py \cite{togo2023first} package for the phonon cross group velocities.}
\label{fig:1-2}
\end{figure}

We thoroughly study the thermal transport of five Ag-based argyrodites, $\rm Ag_{7}PS_{6}$, $\rm Ag_{7}AsS_{6}$, $\rm Ag_{8}SnS_{6}$, $\rm Ag_{8}GeS_{6}$ and $\rm Ag_{9}GaS_{6}$, based on perturbation theory (PT) and the two-channel unified theory (UT) of thermal transport in crystals and glasses including both the population and coherence contributions \cite{simoncelli2019unified}.
As the complexity of the unit cell increases, our results demonstrate that the proportion of population terms decreases while the coherence contributions become more important; thus, $\rm Ag_{7}PS_{6}$ and $\rm Ag_{7}AsS_{6}$ show a fairly weak temperature-dependent $\kappa_{\rm L}$ while the more complex crystalline argyrodites $\rm Ag_{8}SnS_{6}$, $\rm Ag_{8}GeS_{6}$ and $\rm Ag_{9}GaS_{6}$ exhibit a glass-like behavior in their temperature dependence of $\kappa_{\rm L}$, which is ascribed to the strong phonon broadening and the dominance of wavelike phonons. 
Additionally, we observe an ultralow population term of thermal transport ($\kappa_{\rm L}^{\rm p}$), which is attributed to the high phonon scattering rates and the low phonon group velocities. 
Moreover, using laser flash measurements and the homogeneous non-equilibrium molecular dynamics (HNEMD) simulations \cite{fan2019homogeneous,evans1982homogeneous} with our accurate machine learning neuroevolution potentials (NEP) \cite{fan2021neuroevolution}, we also find that the $\kappa_{\rm L}$ of $\rm Ag_{8}SnS_{6}$ and $\rm Ag_{8}GeS_{6}$ exhibit positive temperature dependences and show good consistency with previous measurements \cite{petrov1975characteristics}.
Our study provides a further physical understanding of heat transport in the complex Ag-based argyrodites, facilitating the development of materials that exhibit crystalline, amorphous conductions, or even those with a transition between the two categories.

\section{METHODS}

\subsection{Interatomic force constants}
The initial reference set of $\rm Ag_{8}SnS_{6}$ and $\rm Ag_{8}GeS_{6}$ comprises 250 displaced configurations generated from $ab$ $initio$ molecular dynamics (AIMD) simulations with a 240-atom supercell at temperatures up to 500 K under $NVT$ ensemble, on which single-point DFT calculations are performed. The neuroevolution potential (NEP) was recently proposed \cite{fan2021neuroevolution, fan2022improving} as a promising tool to study heat transport of anharmonic materials such as covalent organic frameworks \cite{li2024active} and silver halide \cite{ouyang2023role} with high accuracy and low computational cost. A radial and an angular cutoff radiuses of 8.0 and 5.0 Å, respectively, are applied to describe the maximum interaction distance between atoms. The fitting weights for atomic energies, forces, and stresses are $\omega_{e}$ = 1, $\omega_{f}$ = 0.01, and $\omega_{s}$ = 0.001, respectively. 10\% of the configurations that are not included in the training set are randomly selected to validate the accuracy and transferability of the NEP. The convergence trends of the root-mean-square errors (RMSEs) of energy and force during the training and testing processes for $\rm Ag_{8}SnS_{6}$ and $\rm Ag_{8}GeS_{6}$ are shown in Figure S2.

\subsection{Extraction of temperature-dependent force constants}
We use the temperature-dependent effective potential (TDEP) scheme as implemented in the hiPhive package \cite{eriksson2019hiphive}, following the work of Hellman \textit{et al.} \cite{hellman2013temperature,hellman2011lattice}, to extract the renormalized second-order force constants. 
AIMD simulations are performed lasting 25 ps with a timestep of 5 fs, and 60 configurations at different temperatures are randomly extracted to fit the temperature-dependent cubic and quartic force constants. 
We subtract the harmonic contributions (0 K) from the total forces before fitting the anharmonic force constants \cite{xia2020microscopic}.  
The neighboring cutoff distances for pairs, triplets, and quadruplets of five argyrodites are 7.0, 5.0, and 4.0 Å, respectively. 
A $q$-mesh of 6×4×3 and a $scalebroad$ of 0.5 are adequate to obtain converged phonon lifetimes of $\rm Ag_{8}SnS_{6}$ and $\rm Ag_{8}GeS_{6}$. 

\subsection{Lattice thermal conductivity calculations}
Lattice thermal conductivities ($\kappa_{\rm L}^{\rm p+c}$) of five argyrodites, $\rm Ag_{7}PS_{6}$, $\rm Ag_{7}AsS_{6}$, $\rm Ag_{8}SnS_{6}$, $\rm Ag_{8}GeS_{6}$ and $\rm Ag_{9}GaS_{6}$ contributed by the diagonal (population $\kappa_{\rm L}^{\rm p}$) and off-diagonal (coherence $\kappa_{\rm L}^{\rm c}$) terms are calculated following the UT proposed by Simoncelli \textit{et al.} \cite{simoncelli2019unified} using the Phono3py \cite{togo2023first} package which solves the Wigner transport equation with the velocity operator in the smooth phase convention of the dynamical matrix. Our in-house scripts are used to convert the three-phonon (3ph) and four-phonon (4ph) scattering rates calculated using the ShengBTE code \cite{han2022fourphonon} to the Phono3py format.
The temperature-dependent $\kappa_{\rm L}$ of $\rm Ag_{8}SnS_{6}$ and $\rm Ag_{8}GeS_{6}$ have also been calculated from the HNEMD simulations \cite{fan2019homogeneous,evans1982homogeneous}. Furthermore, the total temperature-dependent thermal conductivity ($\kappa_{\rm tot}$) is also measured using the laser flash technique. 
Because the band gaps of $\rm Ag_{8}SnS_{6}$ and $\rm Ag_{8}GeS_{6}$ are larger than 1.2 eV, as shown in Table S1 of Supplemental Material (SM) \cite{Supplemental}, the contribution of electronic thermal conductivity to $\kappa_{\rm tot}$ is expected to be small, and $\kappa_{\rm tot}$ is approximately equal to $\kappa_{\rm L}$.

\subsection{Sample preparation, characterization and thermal transport measurement}
Silver (Ag, rods, 99.99\%), germanium (Ge, pellets, 99.9999\%), sulfur (S, sheets, 99.999\%), and tin (Sn, shots, 99.999\%) were used as raw materials. Polycrystalline $\rm Ag_{8}SnS_{6}$ and $\rm Ag_{8}GeS_{6}$ were synthesized by melting the stoichiometric amount of high-purity elements at 1273 K for 24 hours and annealing at 673 K for 48 hours. The obtained ingots were then crushed into fine powders by high-energy ball milling under the protection of argon gas. Finally, all samples were hot-pressed on an induction heating hot press system at 773 K for 5 minutes under a uniaxial pressure of ~50 MPa. The density of dense pellet samples is higher than 95\% of the theoretical value. The average grain sizes for $\rm Ag_{8}SnS_{6}$ and $\rm Ag_{8}GeS_{6}$ are 19.4 and 20.3 nm, respectively, which exceed the main phonon mean free paths in these materials (below 1 nm).

The powder X-ray diffraction (XRD) patterns were collected on the X-ray diffractometer (Rigaku MiniFlex 600-C) using Co $K_{\alpha}$ radiation ($\lambda$ = 1.7890 Å). The room-temperature XRD patterns of $\rm Ag_{8}SnS_{6}$ and $\rm Ag_{8}GeS_{6}$ are shown in Figure S7. The simulated XRD patterns are also presented for comparison.

The thermal diffusivity $\alpha$ was measured on the LFA1000 apparatus (Linseis, Germany) using the laser flash method. The square-shaped samples with an edge length of 12.7 mm and a thickness of 1.5 mm were spray coated with a graphite-based coating Graphit 33 (Kontakt Chemie, Zele, Belgium) to ensure the equal opacity and absorbance of the samples. The thermal conductivity $\kappa$ was determined from $\kappa$ = $\rho$$C_{p}$$\alpha$, where $\rho$ is the volumetric density obtained by the Archimedes method, and $C_{p}$ is the isobaric-specific heat calculated from the Dulong-Petit law.

\begin{figure*}
\centering
\includegraphics[width=1\linewidth]{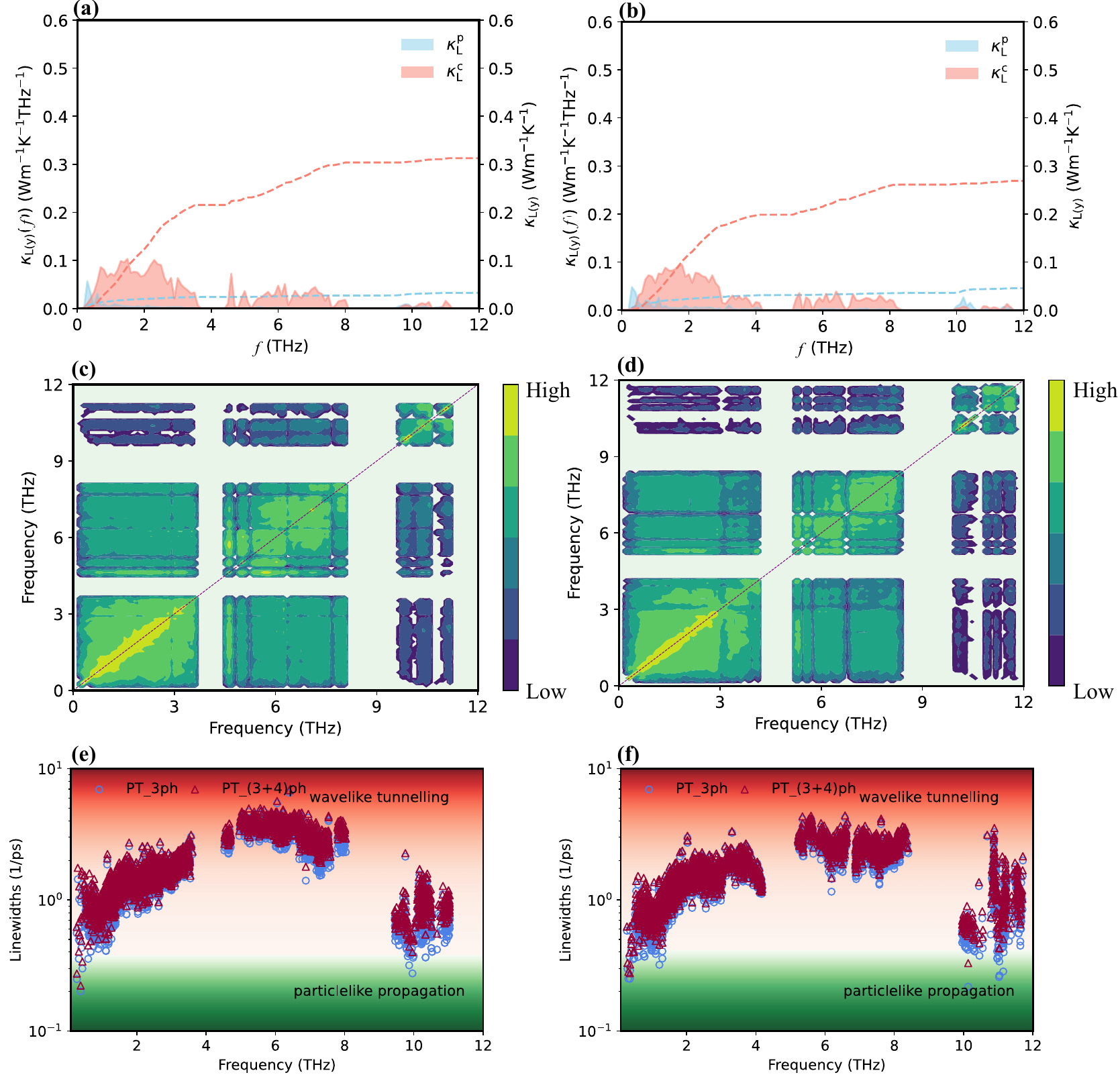}
\caption{\label{fig:wide} Differential and cumulative $\kappa_{\rm L}$ along $y$ direction as a function of the phonon frequency of $\rm Ag_{8}SnS_{6}$ (a) and $\rm Ag_{8}GeS_{6}$ (b) calculated using $\tau_{\rm PT, 3+4}$  within UT at 300 K. The mode-resolved contributions of the coherence term to $\kappa_{\rm L}$ along $y$ direction as a function of various pairs of frequencies for $\rm Ag_{8}SnS_{6}$ (c) and $\rm Ag_{8}GeS_{6}$ (d) at 300 K. The phonon linewidths of $\rm Ag_{8}SnS_{6}$ (e) and $\rm Ag_{8}GeS_{6}$ (f) calculated from PT considering only the third-order anharmonic terms and both third- and fourth-order terms at 300 K. The horizontal line represents the average interband spacing.}
\label{fig:1-3}
\end{figure*}

\section{RESULTS AND DISCUSSION}
Recently,  Mukhopadhyay \textit{et al.} \cite{mukhopadhyay2018two} applied a two-channel model to explain the heat transport mechanism of a promising thermoelectric material $\rm Tl_{3}VSe_{4}$. Their study indicated that the definition of phonons can break down, as heat carriers are not able to transport heat efficiently as particles over long distances compared to normal phonons when the mean free paths ($\Lambda=v\tau$, where $v$ refers to the phonon group velocity, and $\tau$ is the phonon lifetime) are shorter than the minimum interatomic spacing (Ioffe-Rogel limit). We take $\rm Ag_{8}SnS_{6}$ and $\rm Ag_{8}GeS_{6}$ as examples, and their $\Lambda$ are computed using the temperature-dependent phonon group velocities and the phonon lifetimes ($\tau_{\rm PT, 3+4}$) obtained from PT, including both three- and four-phonon interactions, as displayed in Figure S3. We find that the $\Lambda$ of the majority of the heat carriers are smaller than the Loffe-Regel limit in both $\rm Ag_{8}SnS_{6}$ and $\rm Ag_{8}GeS_{6}$, implying the breakdown of the well-defined particle-like phonon picture adopted by the PGM. Hence, the quantitative calculation of $\kappa_{\rm L}$ for these argyrodites needs to consider the contribution of the mutual coherent channel, which describes the wavelike behavior of phonons \cite{hanus2021thermal}.

Based on the recently developed two-channel thermal transport model \cite{simoncelli2019unified} including the population and coherence contributions, we calculate the temperature-dependent $\kappa_{\rm L}^{\rm p+c}$ of five Ag-based argyrodite materials $\rm Ag_{7}PS_{6}$, $\rm Ag_{7}AsS_{6}$, $\rm Ag_{8}SnS_{6}$, $\rm Ag_{8}GeS_{6}$ and $\rm Ag_{9}GaS_{6}$ using $\tau_{\rm PT, 3+4}$ at different temperatures, as shown in Figure 1 (a). Our results illustrate that the temperature-dependence of $\kappa_{\rm L}$ in $\rm Ag_{7}PS_{6}$ and $\rm Ag_{7}AsS_{6}$ is relatively weak while the more complex argyrodites $\rm Ag_{8}SnS_{6}$, $\rm Ag_{8}GeS_{6}$ and $\rm Ag_{9}GaS_{6}$ display a positive temperature dependence of $\kappa_{\rm L}$ despite their crystalline nature, resembling the thermal transport of amorphous materials. 
It is noteworthy that as the complexity of the unit cell of Ag-based argyrodite increases, the temperature dependence of $\kappa_{\rm L}$ becomes more positively correlated with temperature.
To understand the different thermal transport mechanisms in these Ag-based argyrodite materials, we calculate the degree of anharmonicity $\sigma$ \cite{knoop2020anharmonicity}, which is a dimensionless parameter that quantifies the relative strength of anharmonic forces compared to the total forces in a material at a given temperature $T$. It provides a systematic measure of anharmonicity by analyzing forces acting on atoms during MD simulations. \cite{knoop2020anharmonicity}:
\begin{equation}
\sigma^{\mathrm{A}}(T) \equiv \frac{\sigma\left[F^{\mathrm{A}}\right]_T}{\sigma[F]_T}=\sqrt{\frac{\sum_{I, \alpha}\left\langle\left(F_{I, \alpha}^{\mathrm{A}}\right)^2\right\rangle_T}{\sum_{I, \alpha}\left\langle\left(F_{I, \alpha}\right)^2\right\rangle_T}}
\end{equation}
with the thermodynamic ensemble average $\langle  \rangle_{T}$ of each force component $F_{I, \alpha}$ obtained according to:
\begin{equation}
\left\langle (F_{I, \alpha})^2\right\rangle=\frac{1}{N_t} \sum_{t=1}^{N_t} (F_{I, \alpha})^2(t)
\end{equation}
\(\sigma^A(T)\) measures the standard deviation of the distribution of anharmonic force components \(F_{I,\alpha}^A\) at temperature \(T\) along the \(\alpha\) direction. It is apparent that, as the number of atoms in a primitive cell increases, the anharmonicity $\sigma$ also increases. 
Thus, the more structurally complex argyrodites, $\rm Ag_{8}SnS_{6}$, $\rm Ag_{8}GeS_{6}$ and $\rm Ag_{9}GaS_{6}$, exhibit stronger anharmonicity compared with $\rm Ag_{7}PS_{6}$ and $\rm Ag_{7}AsS_{6}$. This indicates that the wavelike phonons can become more significant due to an increased overlap of the closely spaced phonon branches with larger linewidths.

\begin{figure*}
\centering
\includegraphics[width=1\linewidth]{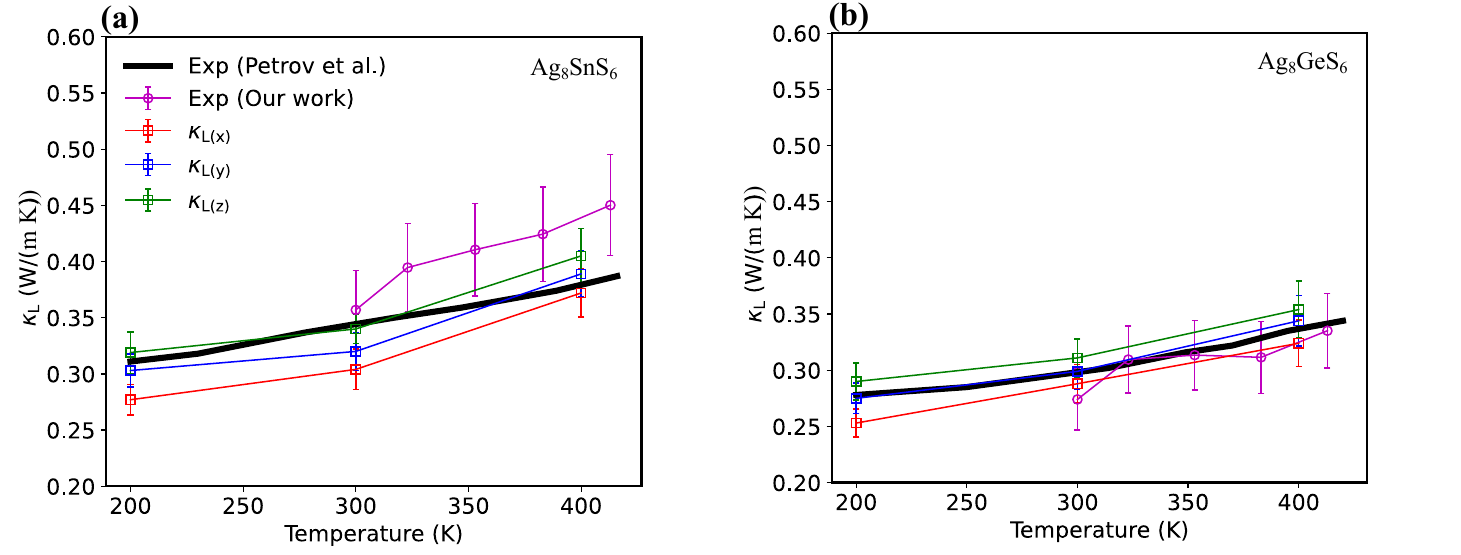}
\caption{\label{fig:wide} $\kappa_{\rm L}$ of $\rm Ag_{8}SnS_{6}$ (a) and $\rm Ag_{8}GeS_{6}$ (b) as a function of temperature calculated using HNEMD simulations along different directions. Our experimental measurements and previous report \cite{petrov1975characteristics} are also shown for comparisons.}
\label{fig:1-4}
\end{figure*}

We further calculate the phonon population and coherence contributions in $\kappa_{\rm L}^{\rm p+c}$ of these Ag-based argyrodites using $\tau_{\rm PT, 3+4}$ at different temperatures, as shown in Figure 2.
Our results demonstrate that, as the complexity of the unit cell of Ag-based argyrodite increases, the $\kappa_{\rm L}^{\rm p}$ declines from 0.141 ($\rm Ag_{7}PS_{6}$) $\rm Wm^{-1}K^{-1}$ and 0.118 ($\rm Ag_{7}AsS_{6}$) to 0.031 ($\rm Ag_{9}GaS_{6}$) $\rm Wm^{-1}K^{-1}$ at 300 K. In contrast, the $\kappa_{\rm L}^{\rm c}$ becomes more dominant; the proportion increases from 70.62\% ($\rm Ag_{7}PS_{6}$) and 71.84\% ($\rm Ag_{7}AsS_{6}$) to 88.39\% ($\rm Ag_{9}GaS_{6}$).
The average $\kappa_{\rm L}^{\rm p}$ of $\rm Ag_{8}SnS_{6}$ and $\rm Ag_{8}GeS_{6}$ are 0.038 and 0.047 $\rm Wm^{-1}K^{-1}$ at 200 K, respectively, accounting for only about 12.50\% and 16.67\% of the total $\kappa_{\rm L}^{\rm p+c}$; the $\kappa_{\rm L}^{\rm c}$ are 0.266 and 0.235 $\rm Wm^{-1}K^{-1}$, accounting for nearly 87.50\% and 83.33\% of the total $\kappa_{\rm L}^{\rm p+c}$. As temperature increases to 400 K, the increased anharmonicity suppresses the average $\kappa_{\rm L}^{\rm p}$ to 0.022 (6.31\%) and 0.026 (7.90\%) $\rm Wm^{-1}K^{-1}$. Contrarily, the coherence terms $\kappa_{\rm L}^{\rm c}$ increase to 0.327 (93.69\%) and 0.303 (92.10\%) $\rm Wm^{-1}K^{-1}$, attributing to the enhanced heat conduction at elevated temperatures. Furthermore, we calculated the phonon population and coherence contributions to the thermal transports of $\rm Ag_{7}PS_{6}$ and $\rm Ag_{7}AsS_{6}$ at different temperatures, where the coherence term also dominates the heat transports but its proportion is relatively smaller, as illustrated in Figure S6.
These results demonstrate the dominating $\kappa_{\rm L}^{\rm c}$ in the more complex crystalline argyrodites $\rm Ag_{8}SnS_{6}$, $\rm Ag_{8}GeS_{6}$ and $\rm Ag_{9}GaS_{6}$ due to the stronger anharmonicity, leading to the positive temperature dependence of the total $\kappa_{\rm L}^{\rm p+c}$.

To further reveal the dominant heat transfer mechanism that originates from the coherence terms in $\rm Ag_{8}SnS_{6}$ and $\rm Ag_{8}GeS_{6}$, we calculate the renormalized phonon dispersion at 300 K (see Figure S1).
The low-lying phonon branches below 1.0 THz and the many flat optical phonon branches result in low group velocities, as demonstrated in Figure S4. The low-lying optical phonons, mainly dominated by Ag-associated vibrations, significantly enhance the scattering rates between acoustic and low-frequency optical modes, leading to large phonon-phonon scattering rates at about 1 THz, as shown in Figure 3 (e, f). Consequently, the combination of the low group velocities and enhanced phonon linewidths leads to an ultralow $\kappa_{\rm L}^{\rm p}$ in $\rm Ag_{8}SnS_{6}$ and $\rm Ag_{8}GeS_{6}$.
However, as a result of their large unit cell (60 atoms in a primitive cell), $\rm Ag_{8}SnS_{6}$ and $\rm Ag_{8}GeS_{6}$ exhibit 177 optical modes and their phonon bands easily overlap. Since the majority of these phonon modes have large phonon linewidths that enhance their tendency to collide with one another, the mutual coherent channels play a crucial role in contributing to the thermal transport.

We also investigate the spectral $\kappa_{\rm L}^{\rm p}$ and $\kappa_{\rm L}^{\rm c}$ and the frequency distribution of the coherence contribution for $\rm Ag_{8}SnS_{6}$ and $\rm Ag_{8}GeS_{6}$ along the $y$ direction, as shown in Figure 3 (a-d). We find that the particle-like phonon channel is mainly contributed by the low-frequency region (0-1 THz), while the wavelike channel is mainly contributed by the high-frequency optical phonons (1-8 THz). The mode pairs away from the diagonal of the frequency plane show large contributions to the diffusive thermal transport in $\rm Ag_{8}SnS_{6}$ and $\rm Ag_{8}GeS_{6}$, which is ascribed to the high phonon-phonon scattering rates and the strongly overlapping phonon branches. We then compute the modal phonon linewidths of propagatons and diffusons \cite{beltukov2013ioffe, simoncelli2022wigner}. According to the Ioffe-Regel criterion in time domain \cite{simoncelli2022wigner}, if the phonon linewidths are larger than the average phonon interband spacing ($\Delta\omega_{\rm ave}$), which is defined as:
\begin{equation}
\begin{aligned}
\Delta\omega_{\rm ave}=\frac{\omega_{\rm max}}{3N}
\end{aligned}
\end{equation}
where $\omega_{\rm max}$ is the maximum phonon frequency and $3N$ represents the number of phonon bands, the corresponding phonon modes are classified as diffusons. Generally, the ratio between the wavelike and particlelike contributions is approximately equivalent to \cite{simoncelli2020generalization}:
\begin{equation}
\frac{{\kappa}_\mathrm{L}^{\mathrm{c(ave)}}(\boldsymbol{q})_s}{{\kappa}_\mathrm{L}^{\mathrm{p(ave)}}(\boldsymbol{q})_s} \simeq \frac{\left[\Delta \omega_{\mathrm{ave}}\right]^{-1}}{\tau(\boldsymbol{q})_s}
\end{equation}
where \(\tau(q)_i = [\Gamma(q)_i]^{-1}\) \cite{sheng1994phonon}, which can be obtained based on the perturbation theory. Therefore, we computed the modal phonon linewidths in $\rm Ag_{8}SnS_{6}$ and $\rm Ag_{8}GeS_{6}$, as displayed in Figure 3 (e, f). It can be seen that the linewidths of the overwhelming majority of phonon modes are larger than $\Delta\omega_{\rm ave}$, demonstrating that the heat in $\rm Ag_{8}SnS_{6}$ and $\rm Ag_{8}GeS_{6}$ is conducted by diffusons, resulting in the observed positive temperature-dependent $\kappa_{\rm L}^{\rm p+c}$.

Our calculated temperature-dependent $\kappa_{\rm L}$ of $\rm Ag_{8}SnS_{6}$ and $\rm Ag_{8}GeS_{6}$ using $\tau_{\rm PT, 3+4}$ within the two-channel Wigner thermal transport model shows a glass-like behavior, which is consistent with our experiments and previous report \cite{petrov1975characteristics}. To better understand the glass-like thermal transport, we further compute the temperature-dependent $\kappa_{\rm L}$ of $\rm Ag_{8}SnS_{6}$ and $\rm Ag_{8}GeS_{6}$ using HNEMD simulations, as displayed in Figure 4 (a-b). 
Furthermore, we calculate the differential and cumulative $\kappa_{\rm L}$ along the $y$ direction as a function of the phonon frequency of $\rm Ag_{8}SnS_{6}$ and $\rm Ag_{8}GeS_{6}$ from the HNEMD simulations with NEP at 300 K, as shown in Figure S10 (a-b). It is obvious that the optical phonons play an important role in the heat transport, which is consistent with our calculations based on the UT framework. Although phonons within the frequencies below 4 THz make a predominant contribution to the heat transport, it is noteworthy that high-frequency phonons (5-11 THz), despite their relatively small group velocities, still play a significant role, indicating that the non-particlelike phonon tunnelling dominates the thermal transport of the high-frequency phonons in $\rm Ag_{8}SnS_{6}$ and $\rm Ag_{8}GeS_{6}$.
The HNEMD results exhibit a good agreement with the experimental data over the temperature range of 200 to 400 K.

\section{CONCLUSION}
In summary, we systematically investigate the thermal transport of five Ag-based argyrodites using the two-channel UT including the population and coherence contributions.
Our results reveal that as the complexity of the unit cell increases, the population terms decline while the coherence contributions become more dominant. As a result, the $\kappa_{\rm L}^{\rm p+c}$ of $\rm Ag_{7}PS_{6}$ and $\rm Ag_{7}AsS_{6}$ show a relatively weak temperature dependence, while more complex crystalline argyrodites $\rm Ag_{8}SnS_{6}$, $\rm Ag_{8}GeS_{6}$ and $\rm Ag_{9}GaS_{6}$ exhibit a glass-like character in their temperature dependence of $\kappa_{\rm L}^{\rm p+c}$, which can be attributed to the strong phonon broadening and the dominance of coherence contributions. 
Additionally, we observe an ultralow $\kappa_{\rm L}^{\rm p}$, which is induced by the strong lattice anharmonicity and the low phonon group velocity. 
Moreover, based on the laser flash measurements and the HNEMD simulations with our machine learning NEP, we provide further evidence for the positive temperature dependence of the $\kappa_{\rm L}$ of $\rm Ag_{8}SnS_{6}$ and $\rm Ag_{8}GeS_{6}$.

\begin{acknowledgments}
This work is supported by the Research Grants Council of Hong Kong (C7002-22Y, 17318122 and 17306721). The authors are grateful for the research computing facilities offered by ITS, HKU.
\end{acknowledgments}

\bibliography{apssamp}

\providecommand{\noopsort}[1]{}\providecommand{\singleletter}[1]{#1}
\begin{thebibliography}{60}%
\makeatletter
\providecommand \@ifxundefined [1]{%
 \@ifx{#1\undefined}
}%
\providecommand \@ifnum [1]{%
 \ifnum #1\expandafter \@firstoftwo
 \else \expandafter \@secondoftwo
 \fi
}%
\providecommand \@ifx [1]{%
 \ifx #1\expandafter \@firstoftwo
 \else \expandafter \@secondoftwo
 \fi
}%
\providecommand \natexlab [1]{#1}%
\providecommand \enquote  [1]{``#1''}%
\providecommand \bibnamefont  [1]{#1}%
\providecommand \bibfnamefont [1]{#1}%
\providecommand \citenamefont [1]{#1}%
\providecommand \href@noop [0]{\@secondoftwo}%
\providecommand \href [0]{\begingroup \@sanitize@url \@href}%
\providecommand \@href[1]{\@@startlink{#1}\@@href}%
\providecommand \@@href[1]{\endgroup#1\@@endlink}%
\providecommand \@sanitize@url [0]{\catcode `\\12\catcode `\$12\catcode `\&12\catcode `\#12\catcode `\^12\catcode `\_12\catcode `\%12\relax}%
\providecommand \@@startlink[1]{}%
\providecommand \@@endlink[0]{}%
\providecommand \url  [0]{\begingroup\@sanitize@url \@url }%
\providecommand \@url [1]{\endgroup\@href {#1}{\urlprefix }}%
\providecommand \urlprefix  [0]{URL }%
\providecommand \Eprint [0]{\href }%
\providecommand \doibase [0]{https://doi.org/}%
\providecommand \selectlanguage [0]{\@gobble}%
\providecommand \bibinfo  [0]{\@secondoftwo}%
\providecommand \bibfield  [0]{\@secondoftwo}%
\providecommand \translation [1]{[#1]}%
\providecommand \BibitemOpen [0]{}%
\providecommand \bibitemStop [0]{}%
\providecommand \bibitemNoStop [0]{.\EOS\space}%
\providecommand \EOS [0]{\spacefactor3000\relax}%
\providecommand \BibitemShut  [1]{\csname bibitem#1\endcsname}%
\let\auto@bib@innerbib\@empty
\bibitem [{\citenamefont {Bell}(2008)}]{bell2008cooling}%
  \BibitemOpen
  \bibfield  {author} {\bibinfo {author} {\bibfnamefont {L.~E.}\ \bibnamefont {Bell}},\ }\bibfield  {title} {\bibinfo {title} {Cooling, heating, generating power, and recovering waste heat with thermoelectric systems},\ }\href@noop {} {\bibfield  {journal} {\bibinfo  {journal} {Science}\ }\textbf {\bibinfo {volume} {321}},\ \bibinfo {pages} {1457} (\bibinfo {year} {2008})}\BibitemShut {NoStop}%
\bibitem [{\citenamefont {Behnia}(2015)}]{behnia2015fundamentals}%
  \BibitemOpen
  \bibfield  {author} {\bibinfo {author} {\bibfnamefont {K.}~\bibnamefont {Behnia}},\ }\href@noop {} {\emph {\bibinfo {title} {Fundamentals of thermoelectricity}}}\ (\bibinfo  {publisher} {OUP Oxford},\ \bibinfo {year} {2015})\BibitemShut {NoStop}%
\bibitem [{\citenamefont {Zhang}\ and\ \citenamefont {Zhao}(2015)}]{zhang2015thermoelectric}%
  \BibitemOpen
  \bibfield  {author} {\bibinfo {author} {\bibfnamefont {X.}~\bibnamefont {Zhang}}\ and\ \bibinfo {author} {\bibfnamefont {L.-D.}\ \bibnamefont {Zhao}},\ }\bibfield  {title} {\bibinfo {title} {Thermoelectric materials: Energy conversion between heat and electricity},\ }\href@noop {} {\bibfield  {journal} {\bibinfo  {journal} {Journal of Materiomics}\ }\textbf {\bibinfo {volume} {1}},\ \bibinfo {pages} {92} (\bibinfo {year} {2015})}\BibitemShut {NoStop}%
\bibitem [{\citenamefont {Humphrey}\ and\ \citenamefont {Linke}(2005)}]{humphrey2005reversible}%
  \BibitemOpen
  \bibfield  {author} {\bibinfo {author} {\bibfnamefont {T.}~\bibnamefont {Humphrey}}\ and\ \bibinfo {author} {\bibfnamefont {H.}~\bibnamefont {Linke}},\ }\bibfield  {title} {\bibinfo {title} {Reversible thermoelectric nanomaterials},\ }\href@noop {} {\bibfield  {journal} {\bibinfo  {journal} {Physical Review Letters}\ }\textbf {\bibinfo {volume} {94}},\ \bibinfo {pages} {096601} (\bibinfo {year} {2005})}\BibitemShut {NoStop}%
\bibitem [{\citenamefont {Snyder}\ and\ \citenamefont {Toberer}(2008)}]{snyder2008complex}%
  \BibitemOpen
  \bibfield  {author} {\bibinfo {author} {\bibfnamefont {G.~J.}\ \bibnamefont {Snyder}}\ and\ \bibinfo {author} {\bibfnamefont {E.~S.}\ \bibnamefont {Toberer}},\ }\bibfield  {title} {\bibinfo {title} {Complex thermoelectric materials},\ }\href@noop {} {\bibfield  {journal} {\bibinfo  {journal} {Nature materials}\ }\textbf {\bibinfo {volume} {7}},\ \bibinfo {pages} {105} (\bibinfo {year} {2008})}\BibitemShut {NoStop}%
\bibitem [{\citenamefont {Sootsman}\ \emph {et~al.}(2009)\citenamefont {Sootsman}, \citenamefont {Chung},\ and\ \citenamefont {Kanatzidis}}]{sootsman2009new}%
  \BibitemOpen
  \bibfield  {author} {\bibinfo {author} {\bibfnamefont {J.~R.}\ \bibnamefont {Sootsman}}, \bibinfo {author} {\bibfnamefont {D.~Y.}\ \bibnamefont {Chung}},\ and\ \bibinfo {author} {\bibfnamefont {M.~G.}\ \bibnamefont {Kanatzidis}},\ }\bibfield  {title} {\bibinfo {title} {New and old concepts in thermoelectric materials},\ }\href@noop {} {\bibfield  {journal} {\bibinfo  {journal} {Angewandte Chemie International Edition}\ }\textbf {\bibinfo {volume} {48}},\ \bibinfo {pages} {8616} (\bibinfo {year} {2009})}\BibitemShut {NoStop}%
\bibitem [{\citenamefont {Tritt}(2011)}]{tritt2011thermoelectric}%
  \BibitemOpen
  \bibfield  {author} {\bibinfo {author} {\bibfnamefont {T.~M.}\ \bibnamefont {Tritt}},\ }\bibfield  {title} {\bibinfo {title} {Thermoelectric phenomena, materials, and applications},\ }\href@noop {} {\bibfield  {journal} {\bibinfo  {journal} {Annual review of materials research}\ }\textbf {\bibinfo {volume} {41}},\ \bibinfo {pages} {433} (\bibinfo {year} {2011})}\BibitemShut {NoStop}%
\bibitem [{\citenamefont {Snyder}\ and\ \citenamefont {Ursell}(2003)}]{snyder2003thermoelectric}%
  \BibitemOpen
  \bibfield  {author} {\bibinfo {author} {\bibfnamefont {G.~J.}\ \bibnamefont {Snyder}}\ and\ \bibinfo {author} {\bibfnamefont {T.~S.}\ \bibnamefont {Ursell}},\ }\bibfield  {title} {\bibinfo {title} {Thermoelectric efficiency and compatibility},\ }\href@noop {} {\bibfield  {journal} {\bibinfo  {journal} {Physical Review Letters}\ }\textbf {\bibinfo {volume} {91}},\ \bibinfo {pages} {148301} (\bibinfo {year} {2003})}\BibitemShut {NoStop}%
\bibitem [{\citenamefont {McGaughey}\ \emph {et~al.}(2019)\citenamefont {McGaughey}, \citenamefont {Jain}, \citenamefont {Kim},\ and\ \citenamefont {Fu}}]{mcgaughey2019phonon}%
  \BibitemOpen
  \bibfield  {author} {\bibinfo {author} {\bibfnamefont {A.~J.}\ \bibnamefont {McGaughey}}, \bibinfo {author} {\bibfnamefont {A.}~\bibnamefont {Jain}}, \bibinfo {author} {\bibfnamefont {H.-Y.}\ \bibnamefont {Kim}},\ and\ \bibinfo {author} {\bibfnamefont {B.}~\bibnamefont {Fu}},\ }\bibfield  {title} {\bibinfo {title} {Phonon properties and thermal conductivity from first principles, lattice dynamics, and the boltzmann transport equation},\ }\href@noop {} {\bibfield  {journal} {\bibinfo  {journal} {Journal of Applied Physics}\ }\textbf {\bibinfo {volume} {125}},\ \bibinfo {pages} {011101} (\bibinfo {year} {2019})}\BibitemShut {NoStop}%
\bibitem [{\citenamefont {Dong}\ \emph {et~al.}(2014)\citenamefont {Dong}, \citenamefont {Cao},\ and\ \citenamefont {Guo}}]{dong2014size}%
  \BibitemOpen
  \bibfield  {author} {\bibinfo {author} {\bibfnamefont {Y.}~\bibnamefont {Dong}}, \bibinfo {author} {\bibfnamefont {B.-Y.}\ \bibnamefont {Cao}},\ and\ \bibinfo {author} {\bibfnamefont {Z.-Y.}\ \bibnamefont {Guo}},\ }\bibfield  {title} {\bibinfo {title} {Size dependent thermal conductivity of {Si} nanosystems based on phonon gas dynamics},\ }\href@noop {} {\bibfield  {journal} {\bibinfo  {journal} {Physica E: Low-dimensional Systems and Nanostructures}\ }\textbf {\bibinfo {volume} {56}},\ \bibinfo {pages} {256} (\bibinfo {year} {2014})}\BibitemShut {NoStop}%
\bibitem [{\citenamefont {Allen}\ \emph {et~al.}(1999)\citenamefont {Allen}, \citenamefont {Feldman}, \citenamefont {Fabian},\ and\ \citenamefont {Wooten}}]{allen1999diffusons}%
  \BibitemOpen
  \bibfield  {author} {\bibinfo {author} {\bibfnamefont {P.~B.}\ \bibnamefont {Allen}}, \bibinfo {author} {\bibfnamefont {J.~L.}\ \bibnamefont {Feldman}}, \bibinfo {author} {\bibfnamefont {J.}~\bibnamefont {Fabian}},\ and\ \bibinfo {author} {\bibfnamefont {F.}~\bibnamefont {Wooten}},\ }\bibfield  {title} {\bibinfo {title} {Diffusons, locons and propagons: Character of atomie yibrations in amorphous {Si}},\ }\href@noop {} {\bibfield  {journal} {\bibinfo  {journal} {Philosophical Magazine B}\ }\textbf {\bibinfo {volume} {79}},\ \bibinfo {pages} {1715} (\bibinfo {year} {1999})}\BibitemShut {NoStop}%
\bibitem [{\citenamefont {Isaeva}\ \emph {et~al.}(2019)\citenamefont {Isaeva}, \citenamefont {Barbalinardo}, \citenamefont {Donadio},\ and\ \citenamefont {Baroni}}]{isaeva2019modeling}%
  \BibitemOpen
  \bibfield  {author} {\bibinfo {author} {\bibfnamefont {L.}~\bibnamefont {Isaeva}}, \bibinfo {author} {\bibfnamefont {G.}~\bibnamefont {Barbalinardo}}, \bibinfo {author} {\bibfnamefont {D.}~\bibnamefont {Donadio}},\ and\ \bibinfo {author} {\bibfnamefont {S.}~\bibnamefont {Baroni}},\ }\bibfield  {title} {\bibinfo {title} {Modeling heat transport in crystals and glasses from a unified lattice-dynamical approach},\ }\href@noop {} {\bibfield  {journal} {\bibinfo  {journal} {Nature Communications}\ }\textbf {\bibinfo {volume} {10}},\ \bibinfo {pages} {3853} (\bibinfo {year} {2019})}\BibitemShut {NoStop}%
\bibitem [{\citenamefont {Seyf}\ \emph {et~al.}(2017)\citenamefont {Seyf}, \citenamefont {Yates}, \citenamefont {Bougher}, \citenamefont {Graham}, \citenamefont {Cola}, \citenamefont {Detchprohm}, \citenamefont {Ji}, \citenamefont {Kim}, \citenamefont {Dupuis}, \citenamefont {Lv} \emph {et~al.}}]{seyf2017rethinking}%
  \BibitemOpen
  \bibfield  {author} {\bibinfo {author} {\bibfnamefont {H.~R.}\ \bibnamefont {Seyf}}, \bibinfo {author} {\bibfnamefont {L.}~\bibnamefont {Yates}}, \bibinfo {author} {\bibfnamefont {T.~L.}\ \bibnamefont {Bougher}}, \bibinfo {author} {\bibfnamefont {S.}~\bibnamefont {Graham}}, \bibinfo {author} {\bibfnamefont {B.~A.}\ \bibnamefont {Cola}}, \bibinfo {author} {\bibfnamefont {T.}~\bibnamefont {Detchprohm}}, \bibinfo {author} {\bibfnamefont {M.-H.}\ \bibnamefont {Ji}}, \bibinfo {author} {\bibfnamefont {J.}~\bibnamefont {Kim}}, \bibinfo {author} {\bibfnamefont {R.}~\bibnamefont {Dupuis}}, \bibinfo {author} {\bibfnamefont {W.}~\bibnamefont {Lv}}, \emph {et~al.},\ }\bibfield  {title} {\bibinfo {title} {Rethinking phonons: The issue of disorder},\ }\href@noop {} {\bibfield  {journal} {\bibinfo  {journal} {npj Computational Materials}\ }\textbf {\bibinfo {volume} {3}},\ \bibinfo {pages} {49} (\bibinfo {year} {2017})}\BibitemShut {NoStop}%
\bibitem [{\citenamefont {Ouyang}\ \emph {et~al.}(2022{\natexlab{a}})\citenamefont {Ouyang}, \citenamefont {Wang},\ and\ \citenamefont {Chen}}]{ouyang2022role}%
  \BibitemOpen
  \bibfield  {author} {\bibinfo {author} {\bibfnamefont {N.}~\bibnamefont {Ouyang}}, \bibinfo {author} {\bibfnamefont {C.}~\bibnamefont {Wang}},\ and\ \bibinfo {author} {\bibfnamefont {Y.}~\bibnamefont {Chen}},\ }\bibfield  {title} {\bibinfo {title} {Role of alloying in the phonon and thermal transport of {SnS--SnSe} across the phase transition},\ }\href@noop {} {\bibfield  {journal} {\bibinfo  {journal} {Materials Today Physics}\ }\textbf {\bibinfo {volume} {28}},\ \bibinfo {pages} {100890} (\bibinfo {year} {2022}{\natexlab{a}})}\BibitemShut {NoStop}%
\bibitem [{\citenamefont {Ouyang}\ \emph {et~al.}(2022{\natexlab{b}})\citenamefont {Ouyang}, \citenamefont {Wang},\ and\ \citenamefont {Chen}}]{ouyang2022temperature}%
  \BibitemOpen
  \bibfield  {author} {\bibinfo {author} {\bibfnamefont {N.}~\bibnamefont {Ouyang}}, \bibinfo {author} {\bibfnamefont {C.}~\bibnamefont {Wang}},\ and\ \bibinfo {author} {\bibfnamefont {Y.}~\bibnamefont {Chen}},\ }\bibfield  {title} {\bibinfo {title} {Temperature-and pressure-dependent phonon transport properties of {SnS} across phase transition from machine-learning interatomic potential},\ }\href@noop {} {\bibfield  {journal} {\bibinfo  {journal} {International Journal of Heat and Mass Transfer}\ }\textbf {\bibinfo {volume} {192}},\ \bibinfo {pages} {122859} (\bibinfo {year} {2022}{\natexlab{b}})}\BibitemShut {NoStop}%
\bibitem [{\citenamefont {Wang}\ \emph {et~al.}(2023)\citenamefont {Wang}, \citenamefont {Zeng}, \citenamefont {Zhao}, \citenamefont {Chen}, \citenamefont {Ouyang}, \citenamefont {Mao},\ and\ \citenamefont {Chen}}]{wang2023b}%
  \BibitemOpen
  \bibfield  {author} {\bibinfo {author} {\bibfnamefont {Q.}~\bibnamefont {Wang}}, \bibinfo {author} {\bibfnamefont {Z.}~\bibnamefont {Zeng}}, \bibinfo {author} {\bibfnamefont {P.}~\bibnamefont {Zhao}}, \bibinfo {author} {\bibfnamefont {C.}~\bibnamefont {Chen}}, \bibinfo {author} {\bibfnamefont {N.}~\bibnamefont {Ouyang}}, \bibinfo {author} {\bibfnamefont {J.}~\bibnamefont {Mao}},\ and\ \bibinfo {author} {\bibfnamefont {Y.}~\bibnamefont {Chen}},\ }\bibfield  {title} {\bibinfo {title} {{B}-site columnar-ordered halide double perovskites: Breaking octahedra motions induces strong lattice anharmonicity and thermal anisotropy},\ }\href@noop {} {\bibfield  {journal} {\bibinfo  {journal} {Chemistry of Materials}\ }\textbf {\bibinfo {volume} {35}},\ \bibinfo {pages} {1633} (\bibinfo {year} {2023})}\BibitemShut {NoStop}%
\bibitem [{\citenamefont {Zheng}\ \emph {et~al.}(2022)\citenamefont {Zheng}, \citenamefont {Shi}, \citenamefont {Yang}, \citenamefont {Lin}, \citenamefont {Huang}, \citenamefont {Guo},\ and\ \citenamefont {Huang}}]{zheng2022anharmonicity}%
  \BibitemOpen
  \bibfield  {author} {\bibinfo {author} {\bibfnamefont {J.}~\bibnamefont {Zheng}}, \bibinfo {author} {\bibfnamefont {D.}~\bibnamefont {Shi}}, \bibinfo {author} {\bibfnamefont {Y.}~\bibnamefont {Yang}}, \bibinfo {author} {\bibfnamefont {C.}~\bibnamefont {Lin}}, \bibinfo {author} {\bibfnamefont {H.}~\bibnamefont {Huang}}, \bibinfo {author} {\bibfnamefont {R.}~\bibnamefont {Guo}},\ and\ \bibinfo {author} {\bibfnamefont {B.}~\bibnamefont {Huang}},\ }\bibfield  {title} {\bibinfo {title} {Anharmonicity-induced phonon hardening and phonon transport enhancement in crystalline perovskite $\rm {BaZrO_{3}}$},\ }\href@noop {} {\bibfield  {journal} {\bibinfo  {journal} {Physical Review B}\ }\textbf {\bibinfo {volume} {105}},\ \bibinfo {pages} {224303} (\bibinfo {year} {2022})}\BibitemShut {NoStop}%
\bibitem [{\citenamefont {Hanus}\ \emph {et~al.}(2021{\natexlab{a}})\citenamefont {Hanus}, \citenamefont {George}, \citenamefont {Wood}, \citenamefont {Bonkowski}, \citenamefont {Cheng}, \citenamefont {Abernathy}, \citenamefont {Manley}, \citenamefont {Hautier}, \citenamefont {Snyder},\ and\ \citenamefont {Hermann}}]{hanus2021uncovering}%
  \BibitemOpen
  \bibfield  {author} {\bibinfo {author} {\bibfnamefont {R.}~\bibnamefont {Hanus}}, \bibinfo {author} {\bibfnamefont {J.}~\bibnamefont {George}}, \bibinfo {author} {\bibfnamefont {M.}~\bibnamefont {Wood}}, \bibinfo {author} {\bibfnamefont {A.}~\bibnamefont {Bonkowski}}, \bibinfo {author} {\bibfnamefont {Y.}~\bibnamefont {Cheng}}, \bibinfo {author} {\bibfnamefont {D.~L.}\ \bibnamefont {Abernathy}}, \bibinfo {author} {\bibfnamefont {M.~E.}\ \bibnamefont {Manley}}, \bibinfo {author} {\bibfnamefont {G.}~\bibnamefont {Hautier}}, \bibinfo {author} {\bibfnamefont {G.~J.}\ \bibnamefont {Snyder}},\ and\ \bibinfo {author} {\bibfnamefont {R.~P.}\ \bibnamefont {Hermann}},\ }\bibfield  {title} {\bibinfo {title} {Uncovering design principles for amorphous-like heat conduction using two-channel lattice dynamics},\ }\href@noop {} {\bibfield  {journal} {\bibinfo  {journal} {Materials Today Physics}\ }\textbf {\bibinfo {volume} {18}},\ \bibinfo {pages} {100344} (\bibinfo {year} {2021}{\natexlab{a}})}\BibitemShut {NoStop}%
\bibitem [{\citenamefont {Wang}\ \emph {et~al.}(2022)\citenamefont {Wang}, \citenamefont {Wang}, \citenamefont {Zhang}, \citenamefont {Chen},\ and\ \citenamefont {Chen}}]{wang2022intrinsic}%
  \BibitemOpen
  \bibfield  {author} {\bibinfo {author} {\bibfnamefont {C.}~\bibnamefont {Wang}}, \bibinfo {author} {\bibfnamefont {Q.}~\bibnamefont {Wang}}, \bibinfo {author} {\bibfnamefont {Q.}~\bibnamefont {Zhang}}, \bibinfo {author} {\bibfnamefont {C.}~\bibnamefont {Chen}},\ and\ \bibinfo {author} {\bibfnamefont {Y.}~\bibnamefont {Chen}},\ }\bibfield  {title} {\bibinfo {title} {Intrinsic {Zn} vacancies-induced wavelike tunneling of phonons and ultralow lattice thermal conductivity in zintl phase $\rm {Sr_{2}ZnSb_{2}}$},\ }\href@noop {} {\bibfield  {journal} {\bibinfo  {journal} {Chemistry of Materials}\ }\textbf {\bibinfo {volume} {34}},\ \bibinfo {pages} {7837} (\bibinfo {year} {2022})}\BibitemShut {NoStop}%
\bibitem [{\citenamefont {Simoncelli}\ \emph {et~al.}(2019)\citenamefont {Simoncelli}, \citenamefont {Marzari},\ and\ \citenamefont {Mauri}}]{simoncelli2019unified}%
  \BibitemOpen
  \bibfield  {author} {\bibinfo {author} {\bibfnamefont {M.}~\bibnamefont {Simoncelli}}, \bibinfo {author} {\bibfnamefont {N.}~\bibnamefont {Marzari}},\ and\ \bibinfo {author} {\bibfnamefont {F.}~\bibnamefont {Mauri}},\ }\bibfield  {title} {\bibinfo {title} {Unified theory of thermal transport in crystals and glasses},\ }\href@noop {} {\bibfield  {journal} {\bibinfo  {journal} {Nature Physics}\ }\textbf {\bibinfo {volume} {15}},\ \bibinfo {pages} {809} (\bibinfo {year} {2019})}\BibitemShut {NoStop}%
\bibitem [{\citenamefont {Luo}\ \emph {et~al.}(2020)\citenamefont {Luo}, \citenamefont {Yang}, \citenamefont {Feng}, \citenamefont {Wang},\ and\ \citenamefont {Ruan}}]{luo2020vibrational}%
  \BibitemOpen
  \bibfield  {author} {\bibinfo {author} {\bibfnamefont {Y.}~\bibnamefont {Luo}}, \bibinfo {author} {\bibfnamefont {X.}~\bibnamefont {Yang}}, \bibinfo {author} {\bibfnamefont {T.}~\bibnamefont {Feng}}, \bibinfo {author} {\bibfnamefont {J.}~\bibnamefont {Wang}},\ and\ \bibinfo {author} {\bibfnamefont {X.}~\bibnamefont {Ruan}},\ }\bibfield  {title} {\bibinfo {title} {Vibrational hierarchy leads to dual-phonon transport in low thermal conductivity crystals},\ }\href@noop {} {\bibfield  {journal} {\bibinfo  {journal} {Nature Communications}\ }\textbf {\bibinfo {volume} {11}},\ \bibinfo {pages} {2554} (\bibinfo {year} {2020})}\BibitemShut {NoStop}%
\bibitem [{\citenamefont {Xia}\ \emph {et~al.}(2020{\natexlab{a}})\citenamefont {Xia}, \citenamefont {Pal}, \citenamefont {He}, \citenamefont {Ozoli{\c{n}}{\v{s}}},\ and\ \citenamefont {Wolverton}}]{xia2020particlelike}%
  \BibitemOpen
  \bibfield  {author} {\bibinfo {author} {\bibfnamefont {Y.}~\bibnamefont {Xia}}, \bibinfo {author} {\bibfnamefont {K.}~\bibnamefont {Pal}}, \bibinfo {author} {\bibfnamefont {J.}~\bibnamefont {He}}, \bibinfo {author} {\bibfnamefont {V.}~\bibnamefont {Ozoli{\c{n}}{\v{s}}}},\ and\ \bibinfo {author} {\bibfnamefont {C.}~\bibnamefont {Wolverton}},\ }\bibfield  {title} {\bibinfo {title} {Particlelike phonon propagation dominates ultralow lattice thermal conductivity in crystalline $\rm {Tl_{3}VSe_{4}}$},\ }\href@noop {} {\bibfield  {journal} {\bibinfo  {journal} {Physical Review Letters}\ }\textbf {\bibinfo {volume} {124}},\ \bibinfo {pages} {065901} (\bibinfo {year} {2020}{\natexlab{a}})}\BibitemShut {NoStop}%
\bibitem [{\citenamefont {Allen}\ and\ \citenamefont {Feldman}(1989)}]{allen1989thermal}%
  \BibitemOpen
  \bibfield  {author} {\bibinfo {author} {\bibfnamefont {P.~B.}\ \bibnamefont {Allen}}\ and\ \bibinfo {author} {\bibfnamefont {J.~L.}\ \bibnamefont {Feldman}},\ }\bibfield  {title} {\bibinfo {title} {Thermal conductivity of glasses: Theory and application to amorphous {Si}},\ }\href@noop {} {\bibfield  {journal} {\bibinfo  {journal} {Physical Review Letters}\ }\textbf {\bibinfo {volume} {62}},\ \bibinfo {pages} {645} (\bibinfo {year} {1989})}\BibitemShut {NoStop}%
\bibitem [{\citenamefont {Sturm}\ \emph {et~al.}(2021)\citenamefont {Sturm}, \citenamefont {Boccalon}, \citenamefont {Ramirez},\ and\ \citenamefont {Kleinke}}]{sturm2021stability}%
  \BibitemOpen
  \bibfield  {author} {\bibinfo {author} {\bibfnamefont {C.}~\bibnamefont {Sturm}}, \bibinfo {author} {\bibfnamefont {N.}~\bibnamefont {Boccalon}}, \bibinfo {author} {\bibfnamefont {D.}~\bibnamefont {Ramirez}},\ and\ \bibinfo {author} {\bibfnamefont {H.}~\bibnamefont {Kleinke}},\ }\bibfield  {title} {\bibinfo {title} {Stability and thermoelectric properties of the canfieldite $\rm {Ag_{8}SnS_{6}}$},\ }\href@noop {} {\bibfield  {journal} {\bibinfo  {journal} {ACS Applied Energy Materials}\ }\textbf {\bibinfo {volume} {4}},\ \bibinfo {pages} {10244} (\bibinfo {year} {2021})}\BibitemShut {NoStop}%
\bibitem [{\citenamefont {Ghrib}\ \emph {et~al.}(2015)\citenamefont {Ghrib}, \citenamefont {Al-Otaibi}, \citenamefont {Almessiere}, \citenamefont {Assaker},\ and\ \citenamefont {Chtourou}}]{ghrib2015high}%
  \BibitemOpen
  \bibfield  {author} {\bibinfo {author} {\bibfnamefont {T.}~\bibnamefont {Ghrib}}, \bibinfo {author} {\bibfnamefont {A.~L.}\ \bibnamefont {Al-Otaibi}}, \bibinfo {author} {\bibfnamefont {M.~A.}\ \bibnamefont {Almessiere}}, \bibinfo {author} {\bibfnamefont {I.~B.}\ \bibnamefont {Assaker}},\ and\ \bibinfo {author} {\bibfnamefont {R.}~\bibnamefont {Chtourou}},\ }\bibfield  {title} {\bibinfo {title} {High thermoelectric figure of merit of $\rm {Ag_{8}SnS_{6}}$ component prepared by electrodeposition technique},\ }\href@noop {} {\bibfield  {journal} {\bibinfo  {journal} {Chinese Physics Letters}\ }\textbf {\bibinfo {volume} {32}},\ \bibinfo {pages} {127402} (\bibinfo {year} {2015})}\BibitemShut {NoStop}%
\bibitem [{\citenamefont {Pogodin}\ \emph {et~al.}(2022)\citenamefont {Pogodin}, \citenamefont {Filep}, \citenamefont {Izai}, \citenamefont {Kokhan},\ and\ \citenamefont {K{\'u}{\v{s}}}}]{pogodin2022crystal}%
  \BibitemOpen
  \bibfield  {author} {\bibinfo {author} {\bibfnamefont {A.}~\bibnamefont {Pogodin}}, \bibinfo {author} {\bibfnamefont {M.}~\bibnamefont {Filep}}, \bibinfo {author} {\bibfnamefont {V.~Y.}\ \bibnamefont {Izai}}, \bibinfo {author} {\bibfnamefont {O.}~\bibnamefont {Kokhan}},\ and\ \bibinfo {author} {\bibfnamefont {P.}~\bibnamefont {K{\'u}{\v{s}}}},\ }\bibfield  {title} {\bibinfo {title} {Crystal growth and electrical conductivity of $\rm {Ag_{7}PS_{6}}$ and $\rm {Ag_{8}GeS_{6}}$ argyrodites},\ }\href@noop {} {\bibfield  {journal} {\bibinfo  {journal} {Journal of Physics and Chemistry of Solids}\ }\textbf {\bibinfo {volume} {168}},\ \bibinfo {pages} {110828} (\bibinfo {year} {2022})}\BibitemShut {NoStop}%
\bibitem [{\citenamefont {Shen}\ \emph {et~al.}(2023)\citenamefont {Shen}, \citenamefont {Koza}, \citenamefont {Tung}, \citenamefont {Ouyang}, \citenamefont {Yang}, \citenamefont {Wang}, \citenamefont {Chen}, \citenamefont {Willa}, \citenamefont {Heid}, \citenamefont {Zhou} \emph {et~al.}}]{shen2023soft}%
  \BibitemOpen
  \bibfield  {author} {\bibinfo {author} {\bibfnamefont {X.}~\bibnamefont {Shen}}, \bibinfo {author} {\bibfnamefont {M.~M.}\ \bibnamefont {Koza}}, \bibinfo {author} {\bibfnamefont {Y.-H.}\ \bibnamefont {Tung}}, \bibinfo {author} {\bibfnamefont {N.}~\bibnamefont {Ouyang}}, \bibinfo {author} {\bibfnamefont {C.-C.}\ \bibnamefont {Yang}}, \bibinfo {author} {\bibfnamefont {C.}~\bibnamefont {Wang}}, \bibinfo {author} {\bibfnamefont {Y.}~\bibnamefont {Chen}}, \bibinfo {author} {\bibfnamefont {K.}~\bibnamefont {Willa}}, \bibinfo {author} {\bibfnamefont {R.}~\bibnamefont {Heid}}, \bibinfo {author} {\bibfnamefont {X.}~\bibnamefont {Zhou}}, \emph {et~al.},\ }\bibfield  {title} {\bibinfo {title} {Soft phonon mode triggering fast ag diffusion in superionic argyrodite $\rm {Ag_{8}GeSe_{6}}$},\ }\href@noop {} {\bibfield  {journal} {\bibinfo  {journal} {Small}\ }\textbf {\bibinfo {volume} {19}},\ \bibinfo {pages} {2305048} (\bibinfo {year} {2023})}\BibitemShut {NoStop}%
\bibitem [{\citenamefont {Carcaly}\ \emph {et~al.}(1979)\citenamefont {Carcaly}, \citenamefont {Ollitrault-Fichet}, \citenamefont {Houphou{\"e}t}, \citenamefont {Eholie},\ and\ \citenamefont {Flahaut}}]{carcaly1979composes}%
  \BibitemOpen
  \bibfield  {author} {\bibinfo {author} {\bibfnamefont {C.}~\bibnamefont {Carcaly}}, \bibinfo {author} {\bibfnamefont {R.}~\bibnamefont {Ollitrault-Fichet}}, \bibinfo {author} {\bibfnamefont {D.}~\bibnamefont {Houphou{\"e}t}}, \bibinfo {author} {\bibfnamefont {R.}~\bibnamefont {Eholie}},\ and\ \bibinfo {author} {\bibfnamefont {J.}~\bibnamefont {Flahaut}},\ }\bibfield  {title} {\bibinfo {title} {Les composes $\rm {Ag_{7}AsS_{6}}$ et $\rm {Ag_{7}AsSe_{6}}$ etude des proprietes thermiques, cristallographiques et electriques},\ }\href@noop {} {\bibfield  {journal} {\bibinfo  {journal} {Materials Research Bulletin}\ }\textbf {\bibinfo {volume} {14}},\ \bibinfo {pages} {1497} (\bibinfo {year} {1979})}\BibitemShut {NoStop}%
\bibitem [{\citenamefont {Petrov}\ \emph {et~al.}(1975)\citenamefont {Petrov}, \citenamefont {Orlov}, \citenamefont {Zaitsev},\ and\ \citenamefont {Feigelman}}]{petrov1975characteristics}%
  \BibitemOpen
  \bibfield  {author} {\bibinfo {author} {\bibfnamefont {A.}~\bibnamefont {Petrov}}, \bibinfo {author} {\bibfnamefont {V.}~\bibnamefont {Orlov}}, \bibinfo {author} {\bibfnamefont {V.}~\bibnamefont {Zaitsev}},\ and\ \bibinfo {author} {\bibfnamefont {V.}~\bibnamefont {Feigelman}},\ }\bibfield  {title} {\bibinfo {title} {Characteristics of the thermal conductivity of $\rm {Ag_{8}MX_{6}}$ compounds having complex crystal structures},\ }\href@noop {} {\bibfield  {journal} {\bibinfo  {journal} {Soviet Physics-Solid State}\ }\textbf {\bibinfo {volume} {17}},\ \bibinfo {pages} {2407} (\bibinfo {year} {1975})}\BibitemShut {NoStop}%
\bibitem [{\citenamefont {Lin}\ \emph {et~al.}(2018)\citenamefont {Lin}, \citenamefont {Li}, \citenamefont {Bu}, \citenamefont {Gao}, \citenamefont {Li},\ and\ \citenamefont {Pei}}]{lin2018thermoelectric}%
  \BibitemOpen
  \bibfield  {author} {\bibinfo {author} {\bibfnamefont {S.}~\bibnamefont {Lin}}, \bibinfo {author} {\bibfnamefont {W.}~\bibnamefont {Li}}, \bibinfo {author} {\bibfnamefont {Z.}~\bibnamefont {Bu}}, \bibinfo {author} {\bibfnamefont {B.}~\bibnamefont {Gao}}, \bibinfo {author} {\bibfnamefont {J.}~\bibnamefont {Li}},\ and\ \bibinfo {author} {\bibfnamefont {Y.}~\bibnamefont {Pei}},\ }\bibfield  {title} {\bibinfo {title} {Thermoelectric properties of $\rm {Ag_{9}GaS_{6}}$ with ultralow lattice thermal conductivity},\ }\href@noop {} {\bibfield  {journal} {\bibinfo  {journal} {Materials Today Physics}\ }\textbf {\bibinfo {volume} {6}},\ \bibinfo {pages} {60} (\bibinfo {year} {2018})}\BibitemShut {NoStop}%
\bibitem [{\citenamefont {Feng}\ \emph {et~al.}(2017)\citenamefont {Feng}, \citenamefont {Lindsay},\ and\ \citenamefont {Ruan}}]{feng2017four}%
  \BibitemOpen
  \bibfield  {author} {\bibinfo {author} {\bibfnamefont {T.}~\bibnamefont {Feng}}, \bibinfo {author} {\bibfnamefont {L.}~\bibnamefont {Lindsay}},\ and\ \bibinfo {author} {\bibfnamefont {X.}~\bibnamefont {Ruan}},\ }\bibfield  {title} {\bibinfo {title} {Four-phonon scattering significantly reduces intrinsic thermal conductivity of solids},\ }\href@noop {} {\bibfield  {journal} {\bibinfo  {journal} {Physical Review B}\ }\textbf {\bibinfo {volume} {96}},\ \bibinfo {pages} {161201} (\bibinfo {year} {2017})}\BibitemShut {NoStop}%
\bibitem [{\citenamefont {Togo}(2023)}]{togo2023first}%
  \BibitemOpen
  \bibfield  {author} {\bibinfo {author} {\bibfnamefont {A.}~\bibnamefont {Togo}},\ }\bibfield  {title} {\bibinfo {title} {First-principles phonon calculations with phonopy and phono3py},\ }\href@noop {} {\bibfield  {journal} {\bibinfo  {journal} {Journal of the Physical Society of Japan}\ }\textbf {\bibinfo {volume} {92}},\ \bibinfo {pages} {012001} (\bibinfo {year} {2023})}\BibitemShut {NoStop}%
\bibitem [{\citenamefont {Fan}\ \emph {et~al.}(2019)\citenamefont {Fan}, \citenamefont {Dong}, \citenamefont {Harju},\ and\ \citenamefont {Ala-Nissila}}]{fan2019homogeneous}%
  \BibitemOpen
  \bibfield  {author} {\bibinfo {author} {\bibfnamefont {Z.}~\bibnamefont {Fan}}, \bibinfo {author} {\bibfnamefont {H.}~\bibnamefont {Dong}}, \bibinfo {author} {\bibfnamefont {A.}~\bibnamefont {Harju}},\ and\ \bibinfo {author} {\bibfnamefont {T.}~\bibnamefont {Ala-Nissila}},\ }\bibfield  {title} {\bibinfo {title} {Homogeneous nonequilibrium molecular dynamics method for heat transport and spectral decomposition with many-body potentials},\ }\href@noop {} {\bibfield  {journal} {\bibinfo  {journal} {Physical Review B}\ }\textbf {\bibinfo {volume} {99}},\ \bibinfo {pages} {064308} (\bibinfo {year} {2019})}\BibitemShut {NoStop}%
\bibitem [{\citenamefont {Evans}(1982)}]{evans1982homogeneous}%
  \BibitemOpen
  \bibfield  {author} {\bibinfo {author} {\bibfnamefont {D.~J.}\ \bibnamefont {Evans}},\ }\bibfield  {title} {\bibinfo {title} {Homogeneous {NEMD} algorithm for thermal conductivity—application of non-canonical linear response theory},\ }\href@noop {} {\bibfield  {journal} {\bibinfo  {journal} {Physics Letters A}\ }\textbf {\bibinfo {volume} {91}},\ \bibinfo {pages} {457} (\bibinfo {year} {1982})}\BibitemShut {NoStop}%
\bibitem [{\citenamefont {Fan}\ \emph {et~al.}(2021)\citenamefont {Fan}, \citenamefont {Zeng}, \citenamefont {Zhang}, \citenamefont {Wang}, \citenamefont {Song}, \citenamefont {Dong}, \citenamefont {Chen},\ and\ \citenamefont {Ala-Nissila}}]{fan2021neuroevolution}%
  \BibitemOpen
  \bibfield  {author} {\bibinfo {author} {\bibfnamefont {Z.}~\bibnamefont {Fan}}, \bibinfo {author} {\bibfnamefont {Z.}~\bibnamefont {Zeng}}, \bibinfo {author} {\bibfnamefont {C.}~\bibnamefont {Zhang}}, \bibinfo {author} {\bibfnamefont {Y.}~\bibnamefont {Wang}}, \bibinfo {author} {\bibfnamefont {K.}~\bibnamefont {Song}}, \bibinfo {author} {\bibfnamefont {H.}~\bibnamefont {Dong}}, \bibinfo {author} {\bibfnamefont {Y.}~\bibnamefont {Chen}},\ and\ \bibinfo {author} {\bibfnamefont {T.}~\bibnamefont {Ala-Nissila}},\ }\bibfield  {title} {\bibinfo {title} {Neuroevolution machine learning potentials: Combining high accuracy and low cost in atomistic simulations and application to heat transport},\ }\href@noop {} {\bibfield  {journal} {\bibinfo  {journal} {Physical Review B}\ }\textbf {\bibinfo {volume} {104}},\ \bibinfo {pages} {104309} (\bibinfo {year} {2021})}\BibitemShut {NoStop}%
\bibitem [{\citenamefont {Fan}(2022)}]{fan2022improving}%
  \BibitemOpen
  \bibfield  {author} {\bibinfo {author} {\bibfnamefont {Z.}~\bibnamefont {Fan}},\ }\bibfield  {title} {\bibinfo {title} {Improving the accuracy of the neuroevolution machine learning potential for multi-component systems},\ }\href@noop {} {\bibfield  {journal} {\bibinfo  {journal} {Journal of Physics: Condensed Matter}\ }\textbf {\bibinfo {volume} {34}},\ \bibinfo {pages} {125902} (\bibinfo {year} {2022})}\BibitemShut {NoStop}%
\bibitem [{\citenamefont {Li}\ \emph {et~al.}(2024)\citenamefont {Li}, \citenamefont {Dong}, \citenamefont {Wang}, \citenamefont {Liu},\ and\ \citenamefont {Yang}}]{li2024active}%
  \BibitemOpen
  \bibfield  {author} {\bibinfo {author} {\bibfnamefont {Z.}~\bibnamefont {Li}}, \bibinfo {author} {\bibfnamefont {H.}~\bibnamefont {Dong}}, \bibinfo {author} {\bibfnamefont {J.}~\bibnamefont {Wang}}, \bibinfo {author} {\bibfnamefont {L.}~\bibnamefont {Liu}},\ and\ \bibinfo {author} {\bibfnamefont {J.-Y.}\ \bibnamefont {Yang}},\ }\bibfield  {title} {\bibinfo {title} {Active learning molecular dynamics-assisted insights into ultralow thermal conductivity of two-dimensional covalent organic frameworks},\ }\href@noop {} {\bibfield  {journal} {\bibinfo  {journal} {International Journal of Heat and Mass Transfer}\ }\textbf {\bibinfo {volume} {225}},\ \bibinfo {pages} {125404} (\bibinfo {year} {2024})}\BibitemShut {NoStop}%
\bibitem [{\citenamefont {Ouyang}\ \emph {et~al.}(2023)\citenamefont {Ouyang}, \citenamefont {Zeng}, \citenamefont {Wang}, \citenamefont {Wang},\ and\ \citenamefont {Chen}}]{ouyang2023role}%
  \BibitemOpen
  \bibfield  {author} {\bibinfo {author} {\bibfnamefont {N.}~\bibnamefont {Ouyang}}, \bibinfo {author} {\bibfnamefont {Z.}~\bibnamefont {Zeng}}, \bibinfo {author} {\bibfnamefont {C.}~\bibnamefont {Wang}}, \bibinfo {author} {\bibfnamefont {Q.}~\bibnamefont {Wang}},\ and\ \bibinfo {author} {\bibfnamefont {Y.}~\bibnamefont {Chen}},\ }\bibfield  {title} {\bibinfo {title} {Role of high-order lattice anharmonicity in the phonon thermal transport of silver halide {Ag}{X (X= Cl, Br, I)}},\ }\href@noop {} {\bibfield  {journal} {\bibinfo  {journal} {Physical Review B}\ }\textbf {\bibinfo {volume} {108}},\ \bibinfo {pages} {174302} (\bibinfo {year} {2023})}\BibitemShut {NoStop}%
\bibitem [{\citenamefont {Eriksson}\ \emph {et~al.}(2019)\citenamefont {Eriksson}, \citenamefont {Fransson},\ and\ \citenamefont {Erhart}}]{eriksson2019hiphive}%
  \BibitemOpen
  \bibfield  {author} {\bibinfo {author} {\bibfnamefont {F.}~\bibnamefont {Eriksson}}, \bibinfo {author} {\bibfnamefont {E.}~\bibnamefont {Fransson}},\ and\ \bibinfo {author} {\bibfnamefont {P.}~\bibnamefont {Erhart}},\ }\bibfield  {title} {\bibinfo {title} {The {hiPhive} package for the extraction of high-order force constants by machine learning},\ }\href@noop {} {\bibfield  {journal} {\bibinfo  {journal} {Advanced Theory and Simulations}\ }\textbf {\bibinfo {volume} {2}},\ \bibinfo {pages} {1800184} (\bibinfo {year} {2019})}\BibitemShut {NoStop}%
\bibitem [{\citenamefont {Hellman}\ \emph {et~al.}(2013)\citenamefont {Hellman}, \citenamefont {Steneteg}, \citenamefont {Abrikosov},\ and\ \citenamefont {Simak}}]{hellman2013temperature}%
  \BibitemOpen
  \bibfield  {author} {\bibinfo {author} {\bibfnamefont {O.}~\bibnamefont {Hellman}}, \bibinfo {author} {\bibfnamefont {P.}~\bibnamefont {Steneteg}}, \bibinfo {author} {\bibfnamefont {I.~A.}\ \bibnamefont {Abrikosov}},\ and\ \bibinfo {author} {\bibfnamefont {S.~I.}\ \bibnamefont {Simak}},\ }\bibfield  {title} {\bibinfo {title} {Temperature dependent effective potential method for accurate free energy calculations of solids},\ }\href@noop {} {\bibfield  {journal} {\bibinfo  {journal} {Physical Review B}\ }\textbf {\bibinfo {volume} {87}},\ \bibinfo {pages} {104111} (\bibinfo {year} {2013})}\BibitemShut {NoStop}%
\bibitem [{\citenamefont {Hellman}\ \emph {et~al.}(2011)\citenamefont {Hellman}, \citenamefont {Abrikosov},\ and\ \citenamefont {Simak}}]{hellman2011lattice}%
  \BibitemOpen
  \bibfield  {author} {\bibinfo {author} {\bibfnamefont {O.}~\bibnamefont {Hellman}}, \bibinfo {author} {\bibfnamefont {I.}~\bibnamefont {Abrikosov}},\ and\ \bibinfo {author} {\bibfnamefont {S.}~\bibnamefont {Simak}},\ }\bibfield  {title} {\bibinfo {title} {Lattice dynamics of anharmonic solids from first principles},\ }\href@noop {} {\bibfield  {journal} {\bibinfo  {journal} {Physical Review B}\ }\textbf {\bibinfo {volume} {84}},\ \bibinfo {pages} {180301} (\bibinfo {year} {2011})}\BibitemShut {NoStop}%
\bibitem [{\citenamefont {Xia}\ \emph {et~al.}(2020{\natexlab{b}})\citenamefont {Xia}, \citenamefont {Ozoli{\c{n}}{\v{s}}},\ and\ \citenamefont {Wolverton}}]{xia2020microscopic}%
  \BibitemOpen
  \bibfield  {author} {\bibinfo {author} {\bibfnamefont {Y.}~\bibnamefont {Xia}}, \bibinfo {author} {\bibfnamefont {V.}~\bibnamefont {Ozoli{\c{n}}{\v{s}}}},\ and\ \bibinfo {author} {\bibfnamefont {C.}~\bibnamefont {Wolverton}},\ }\bibfield  {title} {\bibinfo {title} {Microscopic mechanisms of glasslike lattice thermal transport in cubic $\rm {Cu_{12}Sb_{4}S_{13}}$ tetrahedrites},\ }\href@noop {} {\bibfield  {journal} {\bibinfo  {journal} {Physical Review Letters}\ }\textbf {\bibinfo {volume} {125}},\ \bibinfo {pages} {085901} (\bibinfo {year} {2020}{\natexlab{b}})}\BibitemShut {NoStop}%
\bibitem [{\citenamefont {Han}\ \emph {et~al.}(2022)\citenamefont {Han}, \citenamefont {Yang}, \citenamefont {Li}, \citenamefont {Feng},\ and\ \citenamefont {Ruan}}]{han2022fourphonon}%
  \BibitemOpen
  \bibfield  {author} {\bibinfo {author} {\bibfnamefont {Z.}~\bibnamefont {Han}}, \bibinfo {author} {\bibfnamefont {X.}~\bibnamefont {Yang}}, \bibinfo {author} {\bibfnamefont {W.}~\bibnamefont {Li}}, \bibinfo {author} {\bibfnamefont {T.}~\bibnamefont {Feng}},\ and\ \bibinfo {author} {\bibfnamefont {X.}~\bibnamefont {Ruan}},\ }\bibfield  {title} {\bibinfo {title} {Fourphonon: An extension module to shengbte for computing four-phonon scattering rates and thermal conductivity},\ }\href@noop {} {\bibfield  {journal} {\bibinfo  {journal} {Computer Physics Communications}\ }\textbf {\bibinfo {volume} {270}},\ \bibinfo {pages} {108179} (\bibinfo {year} {2022})}\BibitemShut {NoStop}%
\bibitem [{Sup()}]{Supplemental}%
  \BibitemOpen
  \href@noop {} {\bibinfo {title} {See supplemental material at [url will be inserted by publisher] for more computational details.}},\ \bibinfo {note} {this includes Refs.~\cite{kresse1996efficient, blochl1994projector, perdew2008restoring, simoncelli2019unified, boon2018ag, bletskan2017electronic, li2017easily, lin2018thermoelectric, dongre2017comparison, fan2019homogeneous, momma2011vesta, hellman2013temperature, hellman2011lattice, togo2015first}.}\BibitemShut {Stop}%
\bibitem [{\citenamefont {Mukhopadhyay}\ \emph {et~al.}(2018)\citenamefont {Mukhopadhyay}, \citenamefont {Parker}, \citenamefont {Sales}, \citenamefont {Puretzky}, \citenamefont {McGuire},\ and\ \citenamefont {Lindsay}}]{mukhopadhyay2018two}%
  \BibitemOpen
  \bibfield  {author} {\bibinfo {author} {\bibfnamefont {S.}~\bibnamefont {Mukhopadhyay}}, \bibinfo {author} {\bibfnamefont {D.~S.}\ \bibnamefont {Parker}}, \bibinfo {author} {\bibfnamefont {B.~C.}\ \bibnamefont {Sales}}, \bibinfo {author} {\bibfnamefont {A.~A.}\ \bibnamefont {Puretzky}}, \bibinfo {author} {\bibfnamefont {M.~A.}\ \bibnamefont {McGuire}},\ and\ \bibinfo {author} {\bibfnamefont {L.}~\bibnamefont {Lindsay}},\ }\bibfield  {title} {\bibinfo {title} {Two-channel model for ultralow thermal conductivity of crystalline $\rm {Tl_{3}VSe_{4}}$},\ }\href@noop {} {\bibfield  {journal} {\bibinfo  {journal} {Science}\ }\textbf {\bibinfo {volume} {360}},\ \bibinfo {pages} {1455} (\bibinfo {year} {2018})}\BibitemShut {NoStop}%
\bibitem [{\citenamefont {Hanus}\ \emph {et~al.}(2021{\natexlab{b}})\citenamefont {Hanus}, \citenamefont {Gurunathan}, \citenamefont {Lindsay}, \citenamefont {Agne}, \citenamefont {Shi}, \citenamefont {Graham},\ and\ \citenamefont {Jeffrey~Snyder}}]{hanus2021thermal}%
  \BibitemOpen
  \bibfield  {author} {\bibinfo {author} {\bibfnamefont {R.}~\bibnamefont {Hanus}}, \bibinfo {author} {\bibfnamefont {R.}~\bibnamefont {Gurunathan}}, \bibinfo {author} {\bibfnamefont {L.}~\bibnamefont {Lindsay}}, \bibinfo {author} {\bibfnamefont {M.~T.}\ \bibnamefont {Agne}}, \bibinfo {author} {\bibfnamefont {J.}~\bibnamefont {Shi}}, \bibinfo {author} {\bibfnamefont {S.}~\bibnamefont {Graham}},\ and\ \bibinfo {author} {\bibfnamefont {G.}~\bibnamefont {Jeffrey~Snyder}},\ }\bibfield  {title} {\bibinfo {title} {Thermal transport in defective and disordered materials},\ }\href@noop {} {\bibfield  {journal} {\bibinfo  {journal} {Applied Physics Reviews}\ }\textbf {\bibinfo {volume} {8}},\ \bibinfo {pages} {031311} (\bibinfo {year} {2021}{\natexlab{b}})}\BibitemShut {NoStop}%
\bibitem [{\citenamefont {Knoop}\ \emph {et~al.}(2020)\citenamefont {Knoop}, \citenamefont {Purcell}, \citenamefont {Scheffler},\ and\ \citenamefont {Carbogno}}]{knoop2020anharmonicity}%
  \BibitemOpen
  \bibfield  {author} {\bibinfo {author} {\bibfnamefont {F.}~\bibnamefont {Knoop}}, \bibinfo {author} {\bibfnamefont {T.~A.}\ \bibnamefont {Purcell}}, \bibinfo {author} {\bibfnamefont {M.}~\bibnamefont {Scheffler}},\ and\ \bibinfo {author} {\bibfnamefont {C.}~\bibnamefont {Carbogno}},\ }\bibfield  {title} {\bibinfo {title} {Anharmonicity measure for materials},\ }\href@noop {} {\bibfield  {journal} {\bibinfo  {journal} {Physical Review Materials}\ }\textbf {\bibinfo {volume} {4}},\ \bibinfo {pages} {083809} (\bibinfo {year} {2020})}\BibitemShut {NoStop}%
\bibitem [{\citenamefont {Beltukov}\ \emph {et~al.}(2013)\citenamefont {Beltukov}, \citenamefont {Kozub},\ and\ \citenamefont {Parshin}}]{beltukov2013ioffe}%
  \BibitemOpen
  \bibfield  {author} {\bibinfo {author} {\bibfnamefont {Y.}~\bibnamefont {Beltukov}}, \bibinfo {author} {\bibfnamefont {V.}~\bibnamefont {Kozub}},\ and\ \bibinfo {author} {\bibfnamefont {D.}~\bibnamefont {Parshin}},\ }\bibfield  {title} {\bibinfo {title} {Ioffe-regel criterion and diffusion of vibrations in random lattices},\ }\href@noop {} {\bibfield  {journal} {\bibinfo  {journal} {Physical Review B}\ }\textbf {\bibinfo {volume} {87}},\ \bibinfo {pages} {134203} (\bibinfo {year} {2013})}\BibitemShut {NoStop}%
\bibitem [{\citenamefont {Simoncelli}\ \emph {et~al.}(2022)\citenamefont {Simoncelli}, \citenamefont {Marzari},\ and\ \citenamefont {Mauri}}]{simoncelli2022wigner}%
  \BibitemOpen
  \bibfield  {author} {\bibinfo {author} {\bibfnamefont {M.}~\bibnamefont {Simoncelli}}, \bibinfo {author} {\bibfnamefont {N.}~\bibnamefont {Marzari}},\ and\ \bibinfo {author} {\bibfnamefont {F.}~\bibnamefont {Mauri}},\ }\bibfield  {title} {\bibinfo {title} {Wigner formulation of thermal transport in solids},\ }\href@noop {} {\bibfield  {journal} {\bibinfo  {journal} {Physical Review X}\ }\textbf {\bibinfo {volume} {12}},\ \bibinfo {pages} {041011} (\bibinfo {year} {2022})}\BibitemShut {NoStop}%
\bibitem [{\citenamefont {Simoncelli}\ \emph {et~al.}(2020)\citenamefont {Simoncelli}, \citenamefont {Marzari},\ and\ \citenamefont {Cepellotti}}]{simoncelli2020generalization}%
  \BibitemOpen
  \bibfield  {author} {\bibinfo {author} {\bibfnamefont {M.}~\bibnamefont {Simoncelli}}, \bibinfo {author} {\bibfnamefont {N.}~\bibnamefont {Marzari}},\ and\ \bibinfo {author} {\bibfnamefont {A.}~\bibnamefont {Cepellotti}},\ }\bibfield  {title} {\bibinfo {title} {Generalization of fourier’s law into viscous heat equations},\ }\href@noop {} {\bibfield  {journal} {\bibinfo  {journal} {Physical Review X}\ }\textbf {\bibinfo {volume} {10}},\ \bibinfo {pages} {011019} (\bibinfo {year} {2020})}\BibitemShut {NoStop}%
\bibitem [{\citenamefont {Sheng}\ \emph {et~al.}(1994)\citenamefont {Sheng}, \citenamefont {Zhou},\ and\ \citenamefont {Zhang}}]{sheng1994phonon}%
  \BibitemOpen
  \bibfield  {author} {\bibinfo {author} {\bibfnamefont {P.}~\bibnamefont {Sheng}}, \bibinfo {author} {\bibfnamefont {M.}~\bibnamefont {Zhou}},\ and\ \bibinfo {author} {\bibfnamefont {Z.-Q.}\ \bibnamefont {Zhang}},\ }\bibfield  {title} {\bibinfo {title} {Phonon transport in strong-scattering media},\ }\href@noop {} {\bibfield  {journal} {\bibinfo  {journal} {Physical review letters}\ }\textbf {\bibinfo {volume} {72}},\ \bibinfo {pages} {234} (\bibinfo {year} {1994})}\BibitemShut {NoStop}%
\bibitem [{\citenamefont {Kresse}\ and\ \citenamefont {Furthm{\"u}ller}(1996)}]{kresse1996efficient}%
  \BibitemOpen
  \bibfield  {author} {\bibinfo {author} {\bibfnamefont {G.}~\bibnamefont {Kresse}}\ and\ \bibinfo {author} {\bibfnamefont {J.}~\bibnamefont {Furthm{\"u}ller}},\ }\bibfield  {title} {\bibinfo {title} {Efficient iterative schemes for $ab$ $initio$ total-energy calculations using a plane-wave basis set},\ }\href@noop {} {\bibfield  {journal} {\bibinfo  {journal} {Physical Review B}\ }\textbf {\bibinfo {volume} {54}},\ \bibinfo {pages} {11169} (\bibinfo {year} {1996})}\BibitemShut {NoStop}%
\bibitem [{\citenamefont {Bl{\"o}chl}(1994)}]{blochl1994projector}%
  \BibitemOpen
  \bibfield  {author} {\bibinfo {author} {\bibfnamefont {P.~E.}\ \bibnamefont {Bl{\"o}chl}},\ }\bibfield  {title} {\bibinfo {title} {Projector augmented-wave method},\ }\href@noop {} {\bibfield  {journal} {\bibinfo  {journal} {Physical Review B}\ }\textbf {\bibinfo {volume} {50}},\ \bibinfo {pages} {17953} (\bibinfo {year} {1994})}\BibitemShut {NoStop}%
\bibitem [{\citenamefont {Perdew}\ \emph {et~al.}(2008)\citenamefont {Perdew}, \citenamefont {Ruzsinszky}, \citenamefont {Csonka}, \citenamefont {Vydrov}, \citenamefont {Scuseria}, \citenamefont {Constantin}, \citenamefont {Zhou},\ and\ \citenamefont {Burke}}]{perdew2008restoring}%
  \BibitemOpen
  \bibfield  {author} {\bibinfo {author} {\bibfnamefont {J.~P.}\ \bibnamefont {Perdew}}, \bibinfo {author} {\bibfnamefont {A.}~\bibnamefont {Ruzsinszky}}, \bibinfo {author} {\bibfnamefont {G.~I.}\ \bibnamefont {Csonka}}, \bibinfo {author} {\bibfnamefont {O.~A.}\ \bibnamefont {Vydrov}}, \bibinfo {author} {\bibfnamefont {G.~E.}\ \bibnamefont {Scuseria}}, \bibinfo {author} {\bibfnamefont {L.~A.}\ \bibnamefont {Constantin}}, \bibinfo {author} {\bibfnamefont {X.}~\bibnamefont {Zhou}},\ and\ \bibinfo {author} {\bibfnamefont {K.}~\bibnamefont {Burke}},\ }\bibfield  {title} {\bibinfo {title} {Restoring the density-gradient expansion for exchange in solids and surfaces},\ }\href@noop {} {\bibfield  {journal} {\bibinfo  {journal} {Physical Review Letters}\ }\textbf {\bibinfo {volume} {100}},\ \bibinfo {pages} {136406} (\bibinfo {year} {2008})}\BibitemShut {NoStop}%
\bibitem [{\citenamefont {Boon-On}\ \emph {et~al.}(2018)\citenamefont {Boon-On}, \citenamefont {Aragaw}, \citenamefont {Lee}, \citenamefont {Shi},\ and\ \citenamefont {Lee}}]{boon2018ag}%
  \BibitemOpen
  \bibfield  {author} {\bibinfo {author} {\bibfnamefont {P.}~\bibnamefont {Boon-On}}, \bibinfo {author} {\bibfnamefont {B.~A.}\ \bibnamefont {Aragaw}}, \bibinfo {author} {\bibfnamefont {C.-Y.}\ \bibnamefont {Lee}}, \bibinfo {author} {\bibfnamefont {J.-B.}\ \bibnamefont {Shi}},\ and\ \bibinfo {author} {\bibfnamefont {M.-W.}\ \bibnamefont {Lee}},\ }\bibfield  {title} {\bibinfo {title} {$\rm {Ag_{8}SnS_{6}}$: a new {IR} solar absorber material with a near optimal bandgap},\ }\href@noop {} {\bibfield  {journal} {\bibinfo  {journal} {RSC Advances}\ }\textbf {\bibinfo {volume} {8}},\ \bibinfo {pages} {39470} (\bibinfo {year} {2018})}\BibitemShut {NoStop}%
\bibitem [{\citenamefont {Bletskan}\ \emph {et~al.}(2017)\citenamefont {Bletskan}, \citenamefont {Studenyak}, \citenamefont {Vakulchak},\ and\ \citenamefont {Lukach}}]{bletskan2017electronic}%
  \BibitemOpen
  \bibfield  {author} {\bibinfo {author} {\bibfnamefont {D.}~\bibnamefont {Bletskan}}, \bibinfo {author} {\bibfnamefont {I.}~\bibnamefont {Studenyak}}, \bibinfo {author} {\bibfnamefont {V.}~\bibnamefont {Vakulchak}},\ and\ \bibinfo {author} {\bibfnamefont {A.}~\bibnamefont {Lukach}},\ }\bibfield  {title} {\bibinfo {title} {Electronic structure of $\rm {Ag_{8}GeS_{6}}$},\ }\href@noop {} {\bibfield  {journal} {\bibinfo  {journal} {Semiconductor Physics, Quantum Electronics \& Optoelectronics}\ }\textbf {\bibinfo {volume} {20}},\ \bibinfo {pages} {19} (\bibinfo {year} {2017})}\BibitemShut {NoStop}%
\bibitem [{\citenamefont {Li}\ \emph {et~al.}(2017)\citenamefont {Li}, \citenamefont {Liu}, \citenamefont {Zhang}, \citenamefont {Zhang}, \citenamefont {Guo}, \citenamefont {Shen}, \citenamefont {Zhang},\ and\ \citenamefont {Long}}]{li2017easily}%
  \BibitemOpen
  \bibfield  {author} {\bibinfo {author} {\bibfnamefont {Z.}~\bibnamefont {Li}}, \bibinfo {author} {\bibfnamefont {C.}~\bibnamefont {Liu}}, \bibinfo {author} {\bibfnamefont {X.}~\bibnamefont {Zhang}}, \bibinfo {author} {\bibfnamefont {Z.}~\bibnamefont {Zhang}}, \bibinfo {author} {\bibfnamefont {W.}~\bibnamefont {Guo}}, \bibinfo {author} {\bibfnamefont {L.}~\bibnamefont {Shen}}, \bibinfo {author} {\bibfnamefont {L.}~\bibnamefont {Zhang}},\ and\ \bibinfo {author} {\bibfnamefont {Y.}~\bibnamefont {Long}},\ }\bibfield  {title} {\bibinfo {title} {An easily prepared $\rm {Ag_{8}GeS_{6}}$ nanocrystal and its role on the performance enhancement of polymer solar cells},\ }\href@noop {} {\bibfield  {journal} {\bibinfo  {journal} {Organic Electronics}\ }\textbf {\bibinfo {volume} {45}},\ \bibinfo {pages} {247} (\bibinfo {year} {2017})}\BibitemShut {NoStop}%
\bibitem [{\citenamefont {Dongre}\ \emph {et~al.}(2017)\citenamefont {Dongre}, \citenamefont {Wang},\ and\ \citenamefont {Madsen}}]{dongre2017comparison}%
  \BibitemOpen
  \bibfield  {author} {\bibinfo {author} {\bibfnamefont {B.}~\bibnamefont {Dongre}}, \bibinfo {author} {\bibfnamefont {T.}~\bibnamefont {Wang}},\ and\ \bibinfo {author} {\bibfnamefont {G.~K.}\ \bibnamefont {Madsen}},\ }\bibfield  {title} {\bibinfo {title} {Comparison of the green--kubo and homogeneous non-equilibrium molecular dynamics methods for calculating thermal conductivity},\ }\href@noop {} {\bibfield  {journal} {\bibinfo  {journal} {Modelling and Simulation in Materials Science and Engineering}\ }\textbf {\bibinfo {volume} {25}},\ \bibinfo {pages} {054001} (\bibinfo {year} {2017})}\BibitemShut {NoStop}%
\bibitem [{\citenamefont {Momma}\ and\ \citenamefont {Izumi}(2011)}]{momma2011vesta}%
  \BibitemOpen
  \bibfield  {author} {\bibinfo {author} {\bibfnamefont {K.}~\bibnamefont {Momma}}\ and\ \bibinfo {author} {\bibfnamefont {F.}~\bibnamefont {Izumi}},\ }\bibfield  {title} {\bibinfo {title} {{VESTA 3} for three-dimensional visualization of crystal, volumetric and morphology data},\ }\href@noop {} {\bibfield  {journal} {\bibinfo  {journal} {Journal of Applied Crystallography}\ }\textbf {\bibinfo {volume} {44}},\ \bibinfo {pages} {1272} (\bibinfo {year} {2011})}\BibitemShut {NoStop}%
\bibitem [{\citenamefont {Togo}\ and\ \citenamefont {Tanaka}(2015)}]{togo2015first}%
  \BibitemOpen
  \bibfield  {author} {\bibinfo {author} {\bibfnamefont {A.}~\bibnamefont {Togo}}\ and\ \bibinfo {author} {\bibfnamefont {I.}~\bibnamefont {Tanaka}},\ }\bibfield  {title} {\bibinfo {title} {First principles phonon calculations in materials science},\ }\href@noop {} {\bibfield  {journal} {\bibinfo  {journal} {Scripta Materialia}\ }\textbf {\bibinfo {volume} {108}},\ \bibinfo {pages} {1} (\bibinfo {year} {2015})}\BibitemShut {NoStop}%
\end{thebibliography}%

\end{document}